\documentclass{kapproc} % Computer Modern font calls
\usepackage[dvips]{graphicx}

\newcommand{\be}{\begin{equation}}
\newcommand{\ee}{\end{equation}}
\newcommand{\ben}{\begin{eqnarray}}
\newcommand{\een}{\end{eqnarray}}
\newcommand{\bc}{\begin{center}}
\newcommand{\ec}{\end{center}}

\kluwerbib

\begin{document}

\articletitle{Extragalactic gamma-ray sources}

\author{Diego F. Torres}
\affil{Lawrence Livermore Laboratory, 7000 East Ave. L-413,
Livermore, CA 94550} \email{dtorres@igpp.ucllnl.org}

\begin{abstract}
This chapter provides a review of $\gamma$-ray sources lying at
high Galactic latitudes. Their statistical properties and
variability status, as well as studies involving cross
correlations with lower frequency catalogs and multiwavelength
observations, are summarized. The case for active galactic nuclei
is analyzed with special emphasis, since they represent the
largest population of high energy sources known to date. Other
potential $\gamma$-ray emitters (including nearby starburst
galaxies, normal galaxies, molecular clouds in the Galactic halo,
galaxy clusters, and radio galaxies) that may appear in the next
generation of $\gamma$-ray catalogs and, perhaps, that might have
been already observed by EGRET as unidentified detections, are
also discussed.
\end{abstract}

%\begin{keywords}

%\end{keywords}

\section*{Introduction}

The first extragalactic $\gamma$-ray source ever detected was the
quasar 3C273, observed by the COS-B satellite in an active state
(Swanenburg et al. 1978). Since then, many Active Galactic Nuclei
(AGNs) have been detected at high energies. The Third EGRET
Catalog (3EG) lists several tens of sources labelled as AGNs at
various levels of confidence (Hartman et al. 1999), and there are
120 unidentified detections above $|b|>10^{{\rm o}}$. This chapter
provides a discussion on these sources,  the level of confidence
with which it is known that AGNs have been detected,  the physical
mechanisms by which $\gamma$-rays are emitted from active nuclei,
and several related problems.

\section{Low-, mid-, and high-latitude sources}

Gehrels et al. (2000) presented an analysis of the 3EG sources
classified as steady, i.e. those sources having the most
significant catalog detection both during a timescale of years (as
opposed to those having the most significant detection in one
single viewing period), and within 3$\sigma$ of the flux
calculated using the full data set. The objective in cutting the
sample with this criterion is to get rid of those sources that are
most likely flaring and thus, perhaps, AGNs.
%120 sources remain
%after this cut, a number significant enough as to extract
%statistically reliable results.

Plotting the flux of each unidentified source as a function of
their Galactic latitude, a distinction between bright sources at
low latitudes ($|b|<5^{{\rm o}}$) and weak sources at mid
latitudes ($5^{{\rm o}}<|b|<30^{{\rm o}}$) appears. This is
supported by a $\log \,N$--$\log \,S$ plot (see Gehrels et al.
2000),
%see Fig. \ref{log-log},
which shows how different these two samples are. The distinction
is also supported by the different average photon spectral index
(which is 2.18$\pm$0.04 for low-latitude sources, compared to
2.40$\pm$0.04 for mid-latitude sources).
%, for which the
%chance probability that the two averages are obtained from the
%same sample is $10^{-6}$.

There is unambiguous evidence for the existence of a population of
$\gamma$-ray sources at mid latitudes which is a) fainter, b)
softer, and c) has a steeper $\log \,N$--$\log \,S$ distribution
than sources located at lower Galactic heights. The origin of such
a population has been connected with the Gould Belt (e.g., Grenier
1995, 2000, Grenier \& Perrot 2001).
%CH. ENLARGE
The Gould Belt, comprising massive and late type stars, molecular
clouds, and expanding interstellar medium, and located
asymmetrically across the sky, tilted $\sim 20^{{\rm o}}$ across
the Galactic plane, 100--400 pc away from Earth, could provide a
natural scenario for several of the weak sources detected. The
Belt has an enhanced supernova rate, 3 to 5 times higher than the
Galactic value, and the remnants (both diffuse and compact) of
these explosions could constitute the origin of the $\gamma$-ray
emission (see Grenier 2000).
%There is,
%most suggestively, an above-random positional correlation between
%the Gould Belt itself and the mid-latitude sources, which further
%enhance the proposed association, something that GLAST will test
%in detail.

$\log \,N$--$\log \,S$ plots are very useful in exploring features
of high-latitude sources too, and in particular, in comparing the
high-latitude unidentified source distribution with that of AGNs.
The first such attempt was made by \"Ozel and Thompson (1996), who
found that the difference between unidentified and AGN-labelled
$\log \,N$--$\log \,S$ distributions could be understood in terms of
the superposition of an isotropic (extragalactic) and a Galactic
population of sources. Reimer \& Thompson (2001), using the 3EG
compilation, noticed  that care should be taken in distinguishing
between the $\log \,N$--$\log \,S$ plots using the peak and the
average flux of each detection. AGNs extend to higher flux levels
compared to unidentified sources at high galactic latitudes and the
contrast between the average and the peak flux distributions is even
more pronounced for unidentified sources. The latter happens because
sources at high latitude are preferentially identified by their peak
flux only, in some cases rendering the average flux distribution
meaningless. The similarity between the peak-flux distribution for
unidentified high-latitude sources and AGNs might support the case
for the latter being the counterparts of all unidentified
$\gamma$-ray sources. However, as for low Galactic latitudes,
although trends are indicative, a case-by-case analysis is the only
way to judge this fairly.

\section{The case for AGNs}

\subsection{Definitions}

Blazars are AGNs with a) strong flat spectrum radio emission [the
power law index $\alpha > -0.5$, with $S(\nu) \propto \nu^\alpha$]
and/or b) significant optical polarization, and/or c) significant
flux variability in the optical and in other wavelengths. When the
optical variability occurs on short timescales, the objects are
referred to as optically violently variable --OVV-- quasars. The
blazar classification also includes BL Lacertae (BL Lac) objects,
%(actually blazar is short of BL Lac and quasars),
which present a complete or nearly complete lack of emission
lines, and highly polarized quasars (HPQs). It also refers,
sometimes, to flat spectrum radio quasars (FSRQs), although these
are generally more distant, more luminous, and have stronger
emission lines.

Within the unification model, the underlying scenario for all AGNs
is intrinsically similar. At the very center of the galaxy there
is a supermassive black hole ($\sim$10$^6$ to
$\sim$10$^{10}\,$M$_{\odot}$) which accretes galactic matter
forming an accretion disk. Broad emission lines are produced in
clouds orbiting above the disc at high velocity (the Broad Line
Region, BLR), and this central region is surrounded by an
extended, dusty, molecular torus. A hot electron corona populates
the inner region, probably generating continuum X-ray emission.
Narrower emission lines are produced in clouds moving much farther
from the central black hole.  Two-sided jets of relativistic
particles emanate perpendicular to the plane of the accretion
disc, the generation of which is still not fully understood.
Unification of different AGN classes is achieved taken into
account the intrinsic anisotropy of the phenomenon, as shown in
Fig. 1 (see Urry \& Padovani 1995 and Padovani 1997, for a
detailed discussion).

\begin{figure}[t]
\centering
\includegraphics[width=8cm, height=7cm]{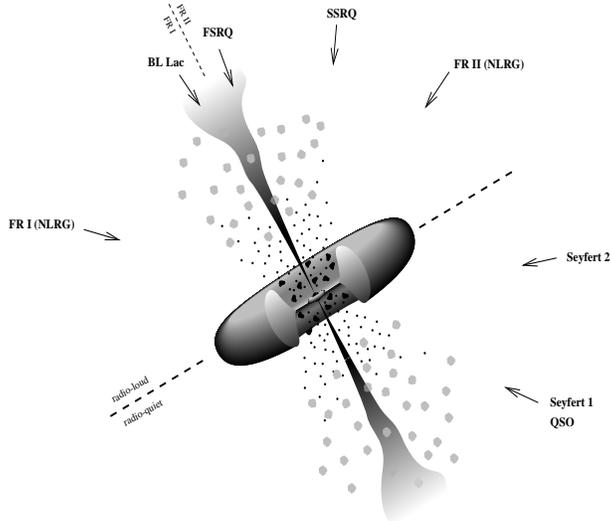}
\caption{The unification model for AGNs. The components of the
figure are discussed in the text. Blazars are those AGNs for which
the jets are close to line of sight. A regular quasar or a Seyfert
1 galaxy is observed if the orientation angle is $\sim 30^{{\rm
o}}$, where the narrow-line and broad-line regions are visible. At
larger angular offsets, the broad-line region will be hidden by
the torus, the corresponding class being Seyfert 2 galaxies.
Perpendicular to the jet axis, the full extent of the jets may be
seen particular at low frequencies, giving rise to a morphology
typical of radio galaxies. The figure is adapted from Urry \&
Padovani (1995) and  Padovani (1997). After Collmar (2001).
\label{uni}}
\end{figure}

\subsection{Gamma-ray emission from blazars}

The fact that some $\gamma$-ray blazars have been observed to
flare dramatically on timescales of days imposes severe
constraints on the size of the emitting region. A direct
constraint can be put on the compactness of the source considering
that the optical depth for $\gamma+\gamma\rightarrow e^+ + e^-$
attenuation is (e.g., Schlickeiser 1996):
\begin{equation}
\tau\simeq\sigma_T n_{\gamma} R= \frac{\sigma_T}{4\pi c
\left<\epsilon\right>} l,
\end{equation}
where $\sigma_T$ is the Thomson cross section, $n_{\gamma}$ is the
$\gamma$-ray photon density, $R<c t_{\rm v}$ is the source size
inferred from the intrinsic variability timescale and simple
light-travel arguments, $\langle\epsilon\rangle$ is the mean photon
energy, and $l=L/R$ is the compactness parameter defined as the
ratio of the intrinsic source luminosity $L$ to its radius.
Considering a mean photon energy of 1 MeV, and (see below)
luminosities of $\sim 10^{48}$ erg s$^{-1}$, the optical depth can
be scaled as $ \tau > 200 {L_{48}}/({t_{\rm v}/{1\;\rm day}}) $. For
fiducial values, the source is opaque to the escape of $\gamma$-ray
photons, contrary to the observed fact that $\gamma$-ray blazars
present a power-law spectrum over several decades of energy (see
below).

On the contrary, if the emission is beamed, special relativistic
effects have to be taken into account. These enter basically
through one quantity, the Doppler factor, $ \delta = \left[\Gamma
(1 - \beta \cos \theta ) \right ]^{-1}, $ where, as usual, $\beta
= V_{\rm jet}/c$ is the velocity of the jet in units of the speed
of light, $\theta$ is the angle between the jet and the line of
sight, and $\Gamma$ is the bulk Lorentz factor
$\Gamma=1/\sqrt{1-\beta^2}$. The Doppler factor regulates the
value of the observed luminosity $L_{\rm app} = \delta ^n L$ (with
$n\sim 3$--4, Begelman et al. 1984), the blueshift of the observed
frequency $\nu_{\rm app} = \delta \nu$, and the time dilation of
events $t_{\rm app}=t/\delta$. Naively, note that when $\theta
\rightarrow 0$, fast jets get Doppler boosted by a large factor,
biasing the detections to closely aligned blazars. Doppler
boosting then makes the compactness factor $l=L/R=\delta^{-1+n}
L_{\rm app}/ct_{\rm app}$, which is then reduced significantly,
such that the optical depth  become less than 1.

The Elliot-Shapiro (1974) relation is a similar argument against
isotropic emission of $\gamma$-rays in the rest frame. For
spherical  accretion onto a black hole, the source luminosity is
limited by the Eddington luminosity
%(what entails the balance between gravitational and radiation forces)
%\begin{equation}
$ L < L_{\rm edd} < 1.3 \times 10^{38}
%\frac
({M/{\rm M}_\odot})/{k}\; {\rm erg\; s}^{-1},
%\end{equation}
$ where the mass of the black hole is given in solar units and $k$
is a correction factor accounting for the difference between
Klein-Nishima and Thomson cross sections (e.g., Dermer and Gehrels
1995, Pohl et al. 1995). If the center of the blazar is a black
hole, the central source has to be larger than the Schwarzschild
radius, implying $\delta t_{v, {\rm min}} > c\,(2GM/c) \sim
10^{-5} (M/$M$_{\odot})$ s. Using the latter two equations, the
relation $ \log {\delta t_{v, {\rm min}}} ({\rm s})
> \log L ({\rm erg\, s}^{-1}) -43 + \log k $,  or, equivalently,
\begin{equation}
\log \left[\frac{\delta t_{{\rm obs}}/(1+z)} {\rm days}\right] >
\log L_{48} + \log k,
\end{equation}
which, for $k$ strictly equal to 1 is known as the Elliot-Shapiro
relation, holds. This inequality is violated for the most rapidly
varying AGNs detected by EGRET. However, this effect, too, is
alleviated if the radiation is beamed, since the reduction in the
compactness parameter $l$ similarly applies here.

In addition, the redshift distribution of detected EGRET sources
(blazars with $z>2$ are part of the sample, see below) shows that
the source distance is not a critical parameter for detection. The
combination of high luminosity, redshift distribution, short-term
variability, and ---as it is shown below--- superluminal motion
observed in blazars, all support the argument for relativistic
beaming in these objects.

\subsection{Models for $\gamma$-ray emission from blazars}

\begin{figure}
\begin{center}
\hspace{1.5cm}\includegraphics[width=2.5cm,height=5cm]{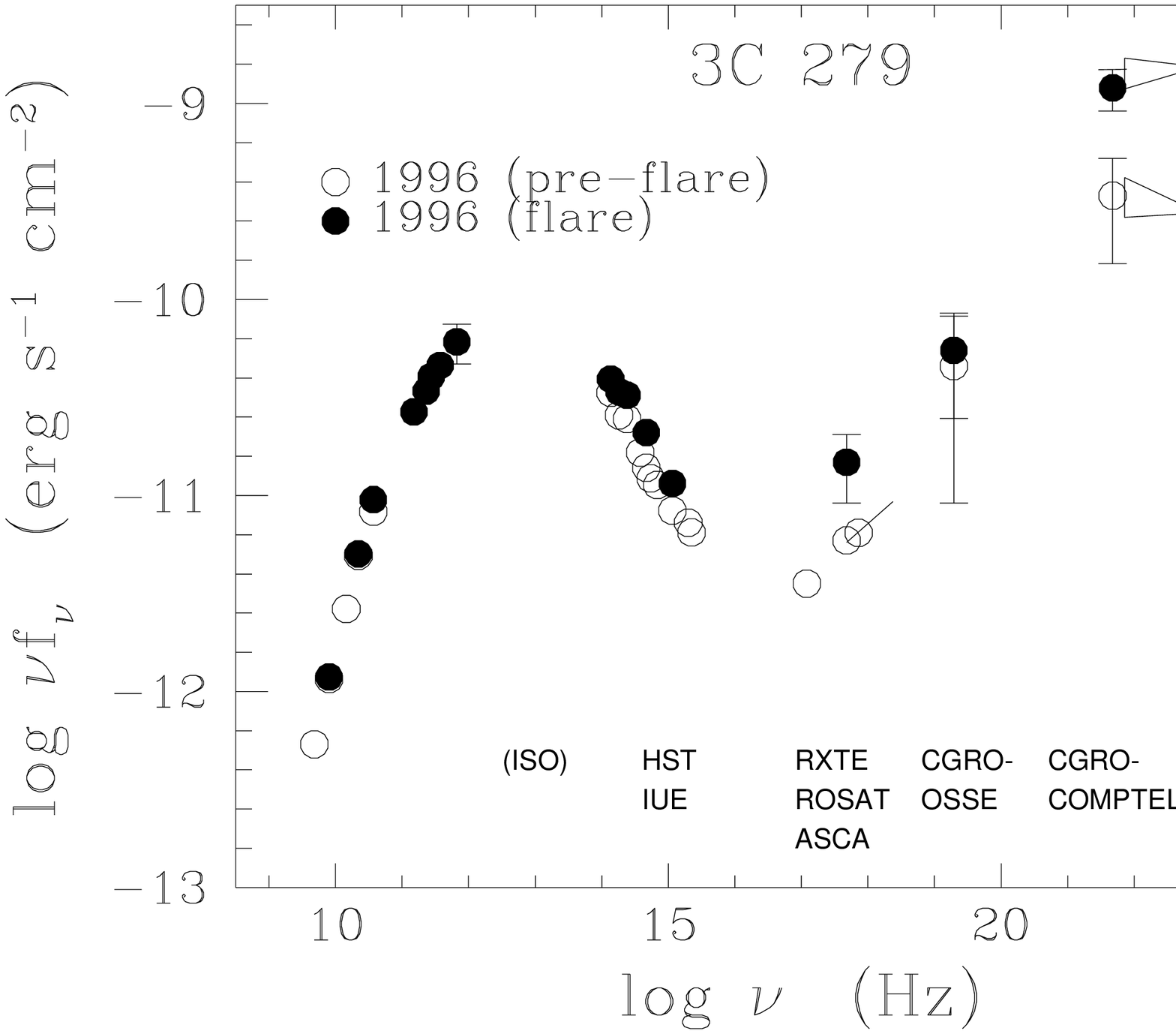}\hspace{1.5cm}
\includegraphics[width=5.5cm,height=4.8cm]{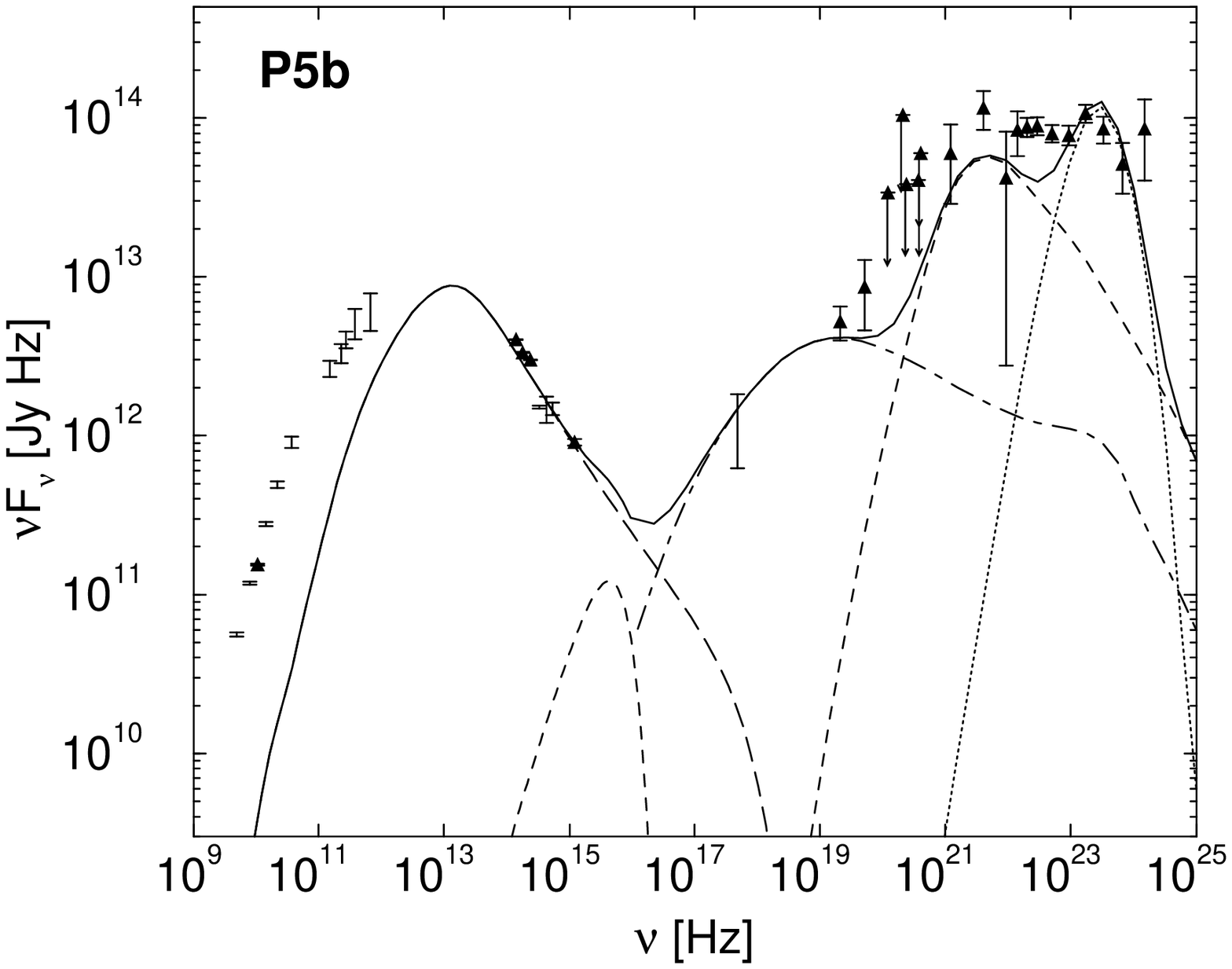}
\end{center}
\caption{Left: Radio to $\gamma$-ray energy distribution of 3C~279
in low (open circles) and high state (filled circles) measured in
January and February, 1996. After Wehrle et al. (1998). Right: Fit
with a leptonic model to the simultaneous broadband spectrum of
3C279 during the time period named P5b, from  January 30 to
February 6, 1996. The low-frequency radio emission is expected to
be produced by less compact regions. Each of the peaks forming the
total curve correspond, respectively, to emission from the
accretion disc, synchrotron, SSC, ECD, and ECC processes. After
Hartman et al. (2001). } \label{3c279}
\end{figure}

Whereas it is generally accepted that blazar emission originates
in relativistic jets, both their nature and the main radiation
mechanism responsible for the observed $\gamma$ radiation is still
under debate. If protons are responsible for the radiation,
photo-pair and photo-pion production, followed by $\pi^0$ decay
and synchrotron emission by secondary particles (e.g., see
Mannheim 1993) would be the main mechanisms by which this
radiation is emitted. If electrons and positrons are the main
constituents of the jets, instead, they would produce
$\gamma$-rays in inverse Compton scattering interactions with the
various seed photon fields traversed by the jet (e.g Marscher \&
Gear 1985, Dermer et al. 1992, Sikora et al. 1994, B\"ottcher et
al. 1997). Rachen (1999) and B\"ottcher (1999) respectively
provide focused reviews on each of these two possibilities. Both
scenarios have to explain the broadband spectra of blazars,
consisting of two components. The first one extends, in the case
of flat-spectrum radio quasars (FSRQs) from radio to optical/UV
frequencies, and in the case of high frequency peaked BL Lacs
(HBLs) up to soft and even hard X-rays.
%, and is consistent with non-thermal synchrotron
%radiation from ultrarelativistic electrons.
The second spectral component appears at $\gamma$-ray energies and
peaks at several MeV to a few GeV in most quasars, or at TeV
energies in some extreme cases. As an example, the spectral energy
distribution of the blazar 3C 279 is shown, for different
observation epochs, in Fig. \ref{3c279} (left panel). It
illustrates  these general features showing a)  strong flux
variability, b)  strong spectral variability, especially when
flaring, and c) the dominance of the $\gamma$-ray emission over
all other wavelengths.

% LEPTONIC

Leptonic jet models explain the radio to UV continuum as
synchrotron radiation from high energy electrons in a
relativistically outflowing jet (e.g., Blandford \& K\"onigl
1979). The emission in the MeV-GeV range is believed to be inverse
Compton scattering of low energy photons by the same relativistic
electrons in the jet. Possible target photon fields for inverse
Compton scattering are the synchrotron photons produced within the
jet (the SSC or Self-Synchrotron Compton process, Marscher \& Gear
1985, Maraschi et al. 1992, Bloom et al. 1996), the UV/soft X-ray
emission from the disk, either entering the jet directly (the ECD
or External Comptonization of Direct disk radiation process,
Dermer et al. 1992, Dermer \& Schlickeiser 1993) or after
reprocessing at the broad-line regions or other circumnuclear
material (the ECC or External Comptonization of radiation from
Clouds, Sikora et al. 1994, Dermer et al. 1997, Blandford and
Levinson 1995), or jet synchrotron radiation reflected at the
broad-line regions (the RS or Reflected Synchrotron mechanism,
Ghisellini \& Madau 1996, B\"ottcher \& Bednarek 1998, Bednarek
1998).

In the SSC process, the relativistic electrons move along the
magnetized jet, generating synchrotron photons with frequency
$\nu_{\rm syn} \propto B\,E^2$, where $B$ is the magnetic field
strength and $E$ is the energy of the particles. These  photons
are upscattered in energy by the same population of electrons that
emitted them, to frequencies $\nu_{\rm IC} \propto \nu_{\rm syn}
\,  E^2 \propto B \, E^4$. To first order, the shape of the IC
spectrum will follow the synchrotron spectrum, which naturally
explains  the spectral turnover in the MeV band, seen in many
blazars.

The same turnover in the spectrum is explained as an
integrated-in-time Compton cooling of the electrons in those
models involving external Compton scattering (EC). The idea is
that when a blob of relativistic electrons is injected into the
jet, and generates an IC spectrum, the high-energy cutoff in that
spectrum (corresponding to the high-energy cutoff of the electron
spectrum) will move towards lower energy as time goes on (higher
energy particles cool first). Integrating the individual spectra
over time generates the spectral turnover at MeV energies (see,
e.g., Hartman et al. 2001 for details on the implementation of the
time average in an individual situation).

Both SSC and EC processes actually occur at the same time, and
their relative importance will depend on the different energy
densities of the target fields. This is the case, for instance, of
PKS 0528+134, which was observed by EGRET from 1991 to 1997
(Mukherjee et al. 1999) and 3C 279, which has simultaneous optical
and X-ray coverage (Hartman et al. 2001, see Fig. \ref{3c279},
right panel). These results show that for high $\gamma$-ray
states, when the ratio between the power of the $\gamma$-ray and
low energy spectral components increased, the bulk Lorentz factor
of the jet increases and the BLR emission Comptonized in the jet
dominates the high-energy spectrum. The SSC mechanism seems to
play a larger role if the blazars are in a low flux state, whereas
for very low activity, even the thermal emission from the disc may
be apparent (see the spectral energy distribution time interval P2
in Hartman et al. 2001, and Pian et al. 1999).

In general, the broadband spectra of HBLs are consistent with pure
SSC models (e.g., Mastiachidis \& Kirk 1997, Pian et al. 1998,
Petry et al. 1999), whereas for FSRQs, a strong contribution of
radiation external to the jet is necessary  to reproduce the
observed spectra. Low-frequency peaked BL Lacs, like BL Lacertae
itself, or W Comae (e.g., B\"ottcher et al. 2002), are expected to
be intermediate cases between radio quasars and HBLs. Ghissellini
et al. (1998) studied this phenomenology analyzing the spectrum of
51 $\gamma$-ray loud quasars for which sufficient data
(non-simultaneous in most cases) were available. They considered
SSC and SSC + EC models and found that there is a well defined
sequence in the properties of HBLs, LBLs, and FSRQs, with
increasing importance of the external radiation field, supporting
a physical (in addition to a geometrical) foundation for the
unification of all quasars.

% HADRONIC

A different flavor of models, so-called hadronic, assume that the
observed $\gamma$-ray emission is initiated by accelerated protons
interacting with either ambient gas or lower frequency radiation.
In the  proton induced cascade (PIC) model (e.g., Mannheim \&
Biermann 1992, Mannheim 1993, Mannheim 1996), the relativistic
protons interact electromagnetically and hadronically by producing
secondary $e^\pm$ pairs (or mesons which eventually decay into
$e^\pm$ pairs), photons, and neutrinos. The observed
$\gamma$-radiation is related to the development of pair cascades
in the jet. The efficiency of this model increases with the energy
of the accelerated protons, usually requiring $E \geq 10^{19} \,
\rm eV$. The synchrotron radiation of protons becomes a very
effective channel of production of high energy $\gamma$-rays at
such energies. Aharonian (2000), Protheroe \& M\"ucke (2000), and
M\"ucke \& Protheroe (2000) have shown that for a reasonable set
of parameters characterizing the small-scale (sub-parsec) jets in
Mrk 421 and Mrk 501, the synchrotron radiation of extremely high
energy protons not only may dominate over other possible radiative
and non-radiative losses, but also could provide adequate fits to
the observed TeV spectra of both objects. This hypothesis could
help to explain the essentially stable spectral shape of Mrk 501,
despite spectacular variations of the TeV flux on sub-day
timescales.

The problem {\it some} hadronic models confront when attempting to
explain TeV flares in blazars is that they require extremely high
luminosities in high energy protons. For instance, for $L_{\rm p}
\leq 10^{45} \, {\rm erg\; s}^{-1}$, the density of the thermal
plasma in the jet exceed  $10^{6} \, \rm cm^{-3}$ in order to
explain the reported TeV flares of Mrk 501 by $\pi^0$-decay
$\gamma$-rays produced by $p$-$p$ interactions (Aharonian 2000).
However, alternative hadronic models assume $\gamma$-ray
production in $pp$ interactions from the collision of jets with
surrounding gas clouds (e.g., Dar \& Laor 1997, Beall \& Bednarek
1999, Purmohammad \& Samimi 2001). Due to the enhanced density in
the cloud,
%(BLR clouds have $n\sim 10^{10}-10^{12}$ cm$^{-3}$),
$pp$ interactions can dominate the $p\gamma$ process, contrary to
the case of the PIC cascade models where $pp$ interactions in the
jet are reduced significantly by the lower jet density. Bednarek \&
Protheroe (1997) proposed that another possible target for the jet
could be the wind of an OB star moving through the jet. Protons have
also been suggested as responsible only for the injection of
energetic electrons, which in turn produce the observed $\gamma$-ray
emission by SSC mechanism (Kazanas \& Mastiachidis 1999).

Correlations of the flux variability at different wavelengths and
on different timescales has been considered as a way to decide
between models. Variability at the highest energies (TeV) ranges
from less than 1~hr (flare on 1996 May 15 of Mrk 421) to long high
states lasting several months (Mrk 501 in 1997); see Aharonian
(1999) and Catanese \& Weekes (1999) for reviews. Multi-wavelength
campaigns revealed that the TeV flares of both, Mrk 501 (see,
e.g., Pian et al. 1998; Catanese et al. 1997; Aharonian et al.
2001; Sambruna et al. 2000) and Mrk 421 (see, e.g., Buckley et al.
1996; Maraschi et al. 1999; Takahashi et al. 1999, also Rebillot
et al. 2003) were correlated with X-ray emission. This fact
appears to favor the synchrotron $+$ Compton jet emission models,
in which the same population of relativistic electrons is
responsible for the production of both X-rays and TeV
$\gamma$-rays.
%via synchrotron radiation and inverse Compton scattering,
%respectively.
However, hadronic models were also shown to be able to reproduce
such correlations as well (see, e.g., Rachen 1999).

Strictly from a `fitting perspective', simultaneous
multiwavelength observations of blazars provide enough elements to
distinguish between leptonic and hadronic models. Such is the case
of W Comae (B\"ottcher et al. 2002) for which the best currently
available contemporaneous optical-to-X-ray spectrum shows clear
evidence for the onset of the high-energy emission component
beyond $\sim 4$ keV. This translates into an accurate indication
of the level of hard X-ray SSC emission in the framework of
leptonic models. B\"ottcher et al. (2002) found that all
acceptable leptonic fits to the optical-to-X-ray emission of W
Comae predict a cutoff of the high-energy emission around $\sim$
100 GeV. The synchrotron-proton blazar model, when fitted to the
same data, predicts similar fluxes at $\sim $ 40 GeV but
noticeable greater fluxes above 100 GeV, at levels reachable by
ground-based telescopes. This contrast may then be used as a
diagnostic of the main emission mechanism.

One unambiguous signature of a hadronic production of
$\gamma$-rays is the concurrent emission of neutrinos. Charge pion
decay, as opposed to neutral pions, will unavoidably produce
energetic neutrinos. The detection of a strong neutrino flux ---in
the next generation km$^2$ neutrino telescopes--- from blazar jets
would identify hadrons as the primary accelerated particles (e.g.,
Nellen et al. 1993, Bednarek \& Protheroe 1999, Stecker et al.
1996, Schuster et al. 2001).

Independently of how the $\gamma$-rays are produced, they must
traverse the strong X-ray field produced in the innermost region
of the accretion disk. The observed $\gamma$-ray photons cannot
originate from too small a radius, otherwise they would be
absorbed through pair creation in the disk photosphere (e.g., %Becker \& Kafatos 1995,
Blandford \& Levinson 1995). This naturally leads to the concept
of $\gamma$-spheres in AGNs: for each $\gamma$-ray photon energy
there is a radius $r_{\gamma}$ beyond which the pair production
opacity to infinity equals unity. The size of the $\gamma$-sphere
depends on both the energy of the $\gamma$-ray photons and the
soft photon flux, as further discussed below.

\section{EGRET Observations of AGNs}

The 3EG catalog distinguishes with an `A' those detections
classified as `high confidence AGNs' and with an `a' those which
are considered lower confidence identifications. A-AGNs were found
within the 95\% confidence location contours, and present a 5 GHz
radio flux in excess of 1 Jy together with a radio spectral index
generally above $\alpha \sim -0.5$. The `a' class is more vaguely
defined: they are near, but outside, the 95\% confidence contour
and the candidate counterpart has a lower radio flux. A
description of the main features of presumed EGRET AGNs follows.
See von Montigny et al. (1995a) and Mukherjee et al. (1997) for
further details.

{\it Power:} The $\gamma$-ray luminosity often dominates the
bolometric power of the blazar.

{\it Redshift and Spectra:} The redshifts of 3EG catalog blazars
range from 0.03 to 2.28. Mukherjee et al. (1997) have noticed that
there are marginal (less than 2.5$\sigma$) indications suggesting
that BL Lacs have slightly harder spectrum in the EGRET energy range
than  FSRQs. Also, for some individual blazars, a trend for the
spectrum to harden during a flare (e.g., in blazars PKS 1222+216,
1633+382, and 0528+134, Sreekumar et al. 1996, Mukherjee et al.
1996) has been noted.

{\it Luminosity:} Given the redshift, the $\gamma$-ray luminosity
can be estimated by considering the relationship between the
observed differential energy flux $S_0(E_0),$ and the power
emitted in $dE$ \be Q_e[E_0]=4\pi S_0(E_0)(1+z)^{\Gamma-1}\Theta
{D_L}^2(z,q_0) \ee where
%\be
$ D_L={c/({H_0{q_0}^2})} \bigl[
1-q_0+q_0z+(q_0-1)(2q_0z+1)^{1/2}\bigr]
%\equiv{{cz}/({H_0})}g(z,q_0),
$
%\ee
with $E=E_0(1+z)$ in the Friedman universe, $H_0$ ($\sim$70 km
s$^{-1}$ Mpc$^{-1}$) is the Hubble parameter, $q_0$ ($\sim$0.5) is
the deceleration parameter today,  $z$ is the redshift, and
$\Theta$ is the beaming factor (Mukherjee et al. 1997).
Luminosities larger than 10$^{45}$ erg s$^{-1}$ are typically
deduced.

{\it Variability:} It is not uncommon, as discussed above, for EGRET
blazars to be detected by their peak flux. Therefore, flux
variations can span more than one order of magnitude between
different observations. PKS 1622$-$297 (Mattox et al. 1997b), for
instance, had a doubling time, i.e. the times during which the
source flux doubled, of less than 3.8 hr. For 3C 279 (Wehrle et al.
1998), the doubling time was $\sim 8$ hr.
%Flux variations over the period
%of a few (1--3) days were detected for 3C 279 (Kniffen et al.
%1993, Wehrle et al. 1998), 3C 454.3 (Hartman et al. 1993), CTA 26
%(Mattox et al. 1995, Hallum et al. 1997, Mattox et al. 2000), 4C
%38.41 (Mattox et al. 1993), PKS 1406-076 (Wagner et al. 1995), and
%PKS 0528+134 (Hunter et al. 1993; Mukherjee et al. 1996).
Wallace et al. (2000) presented a systematic search for short-term
variability
%: the standard method of EGRET data analysis, maximum
%likelihood, was applied to short-duration maps of $\gamma$-ray
%intensity
(sensitive to variations in 1 to 2 days); six 3EG catalog sources
were found to exhibit variability in this study (PKS 0528+134, 3C
66A, and four other unidentified detections).

\subsection{Comparison of properties of AGNs and unidentified EGRET
sources at high latitudes}

Fig. 3 shows the distribution of the $\gamma$-ray photon spectral
index for 45 unidentified EGRET sources above $|b|>30^{{\rm o}}$
and the 66 3EG catalog AGNs. The mean value of the photon index is
2.36$\pm$0.36 for AGNs and 2.49$\pm$0.34 for the unidentified
detections. These are compatible within the uncertainties and, on
average, steeper than what is observed for low-latitude sources.
\begin{figure*}[t]
\begin{center}
\includegraphics[width=5cm,height=6cm]{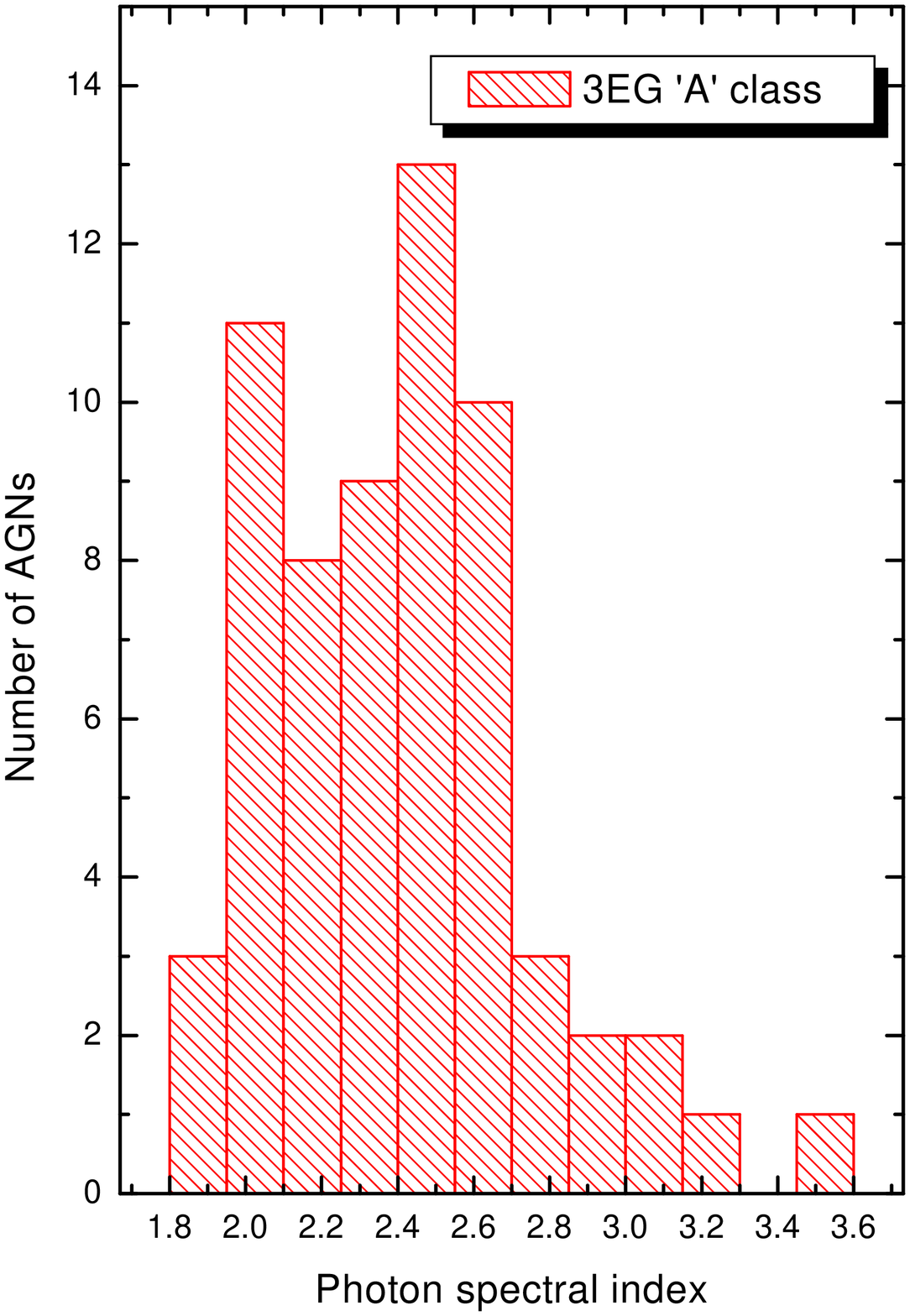}
\includegraphics[width=5cm,height=6cm]{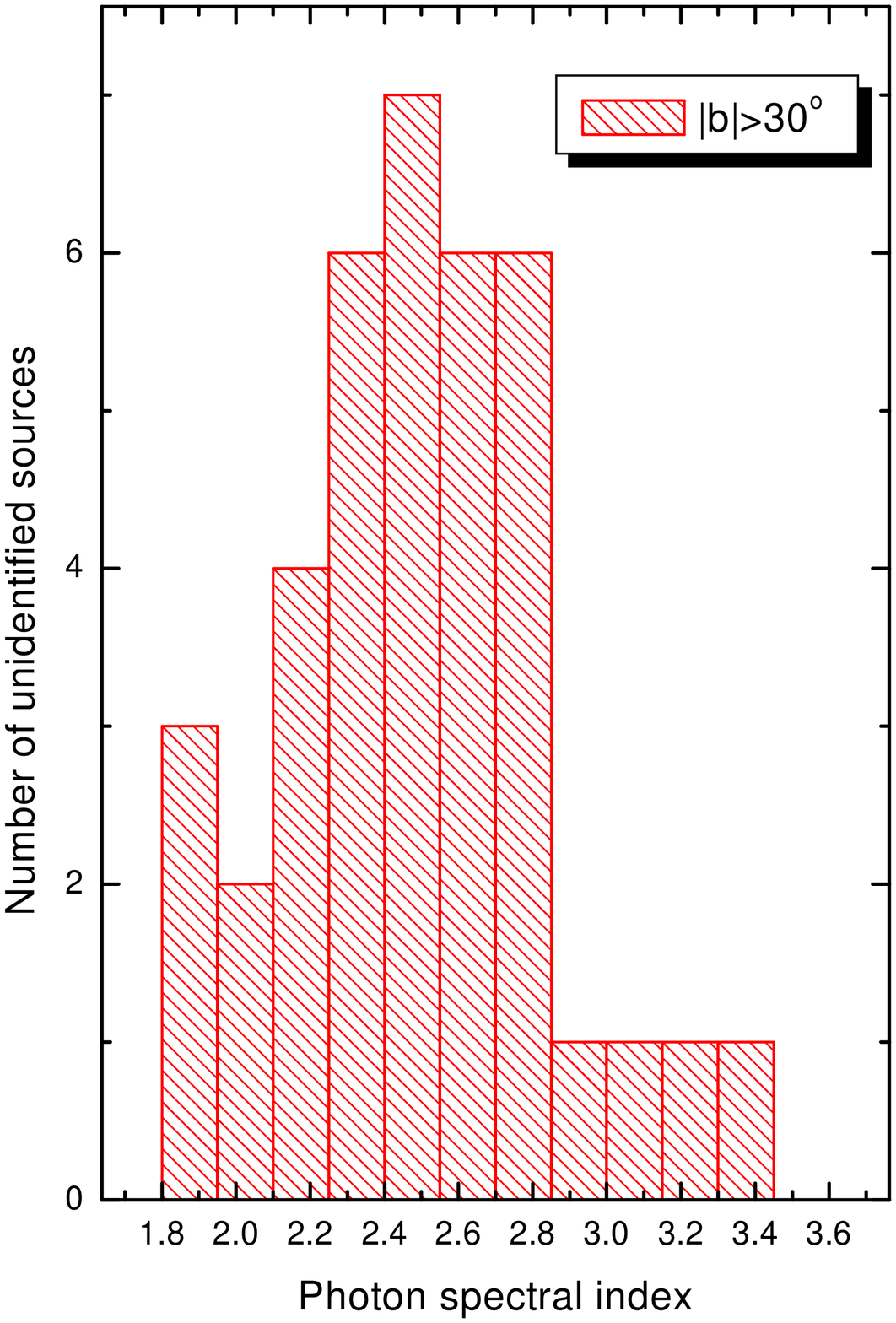}\vspace{-.6cm}
\includegraphics[width=5cm,height=6cm]{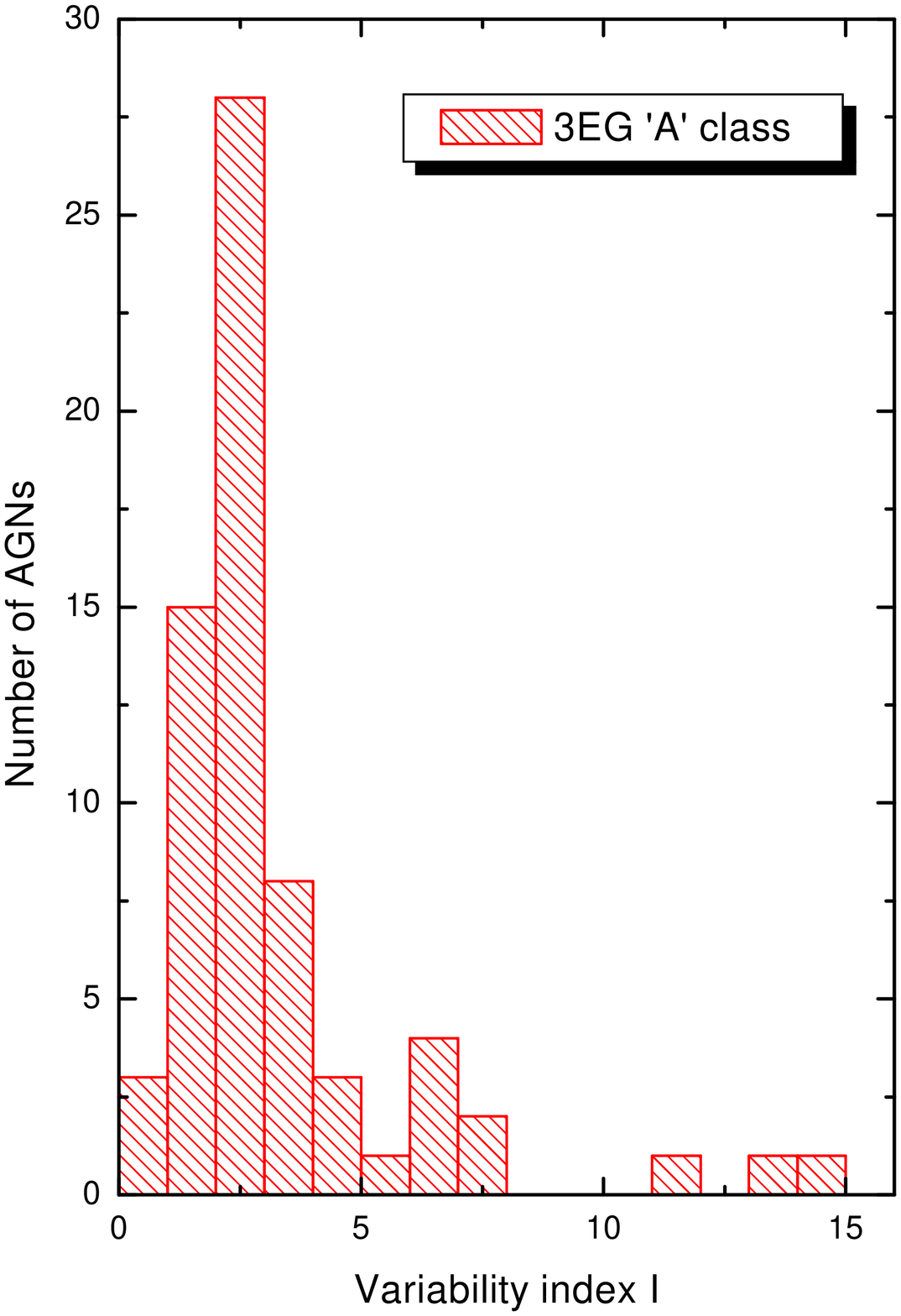}
\includegraphics[width=5cm,height=6cm]{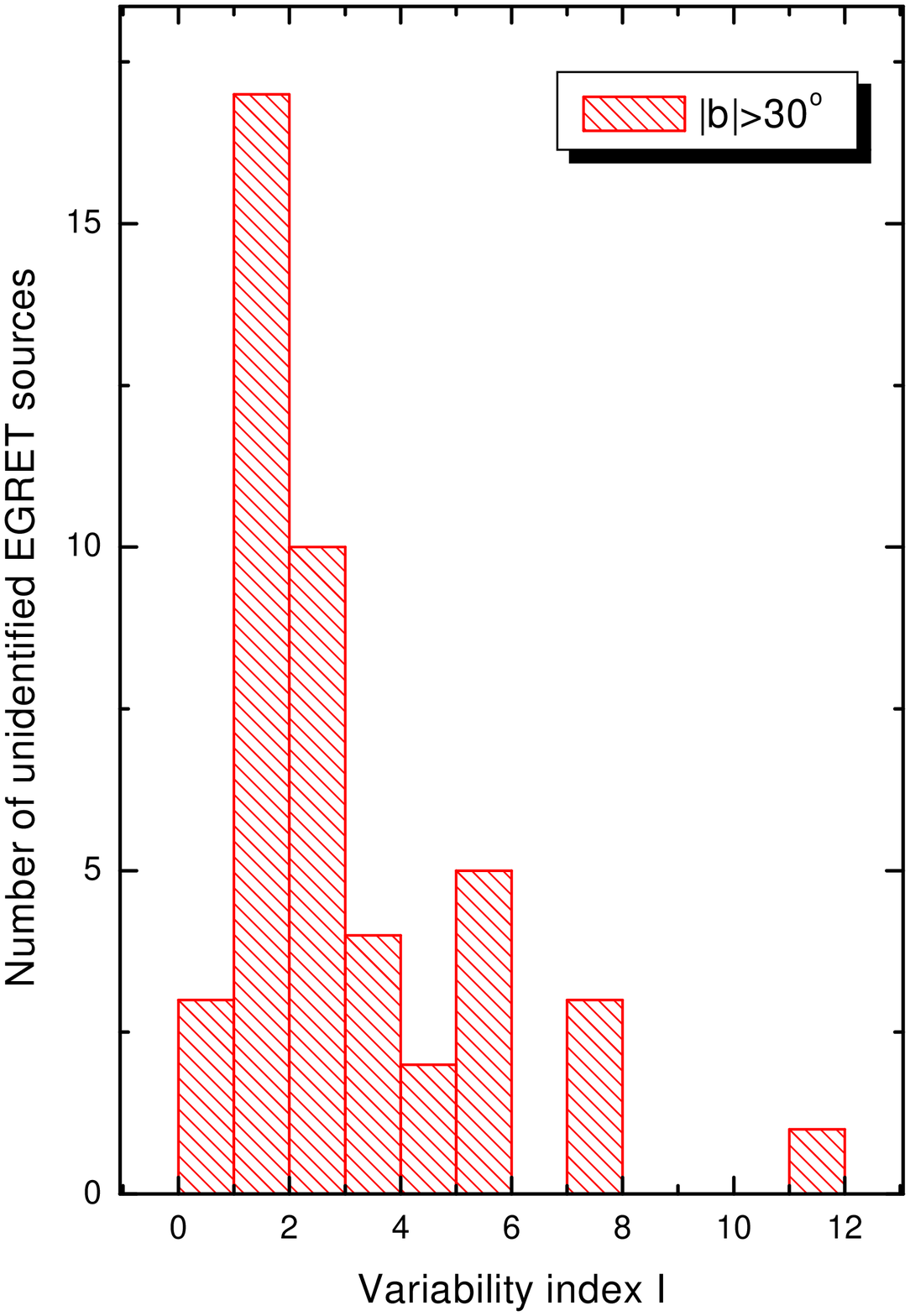}
\caption{Top panels: Photon spectral index distribution.  Bottom
panels: Variability index $I$ distribution. The left panel shows,
in both cases, the distribution for the 66 detected A-AGNs.
Similarly, the right panel shows the corresponding distribution
for the $|b|>30^{\rm o}$ unidentified sources. After Torres et al.
(2003).}\end{center} \label{spec2}
\end{figure*}
Fig. 3 also shows the variability distributions, under the
$I$-index (Torres et al. 2001a) ---use of Tompkins' (1999) index
$\tau$ (see also Nolan et al. 2003) would provide statistically
correlated results (see Torres et al. 2001b). The mean value for
AGNs (lower left panel) is 3.3$\pm$2.6. A peak in the plot is seen
at $I=2.5$, which represents a value $4\sigma$ above that shown by
pulsars. Clearly, most of the AGNs are likely variable sources.
The mean for the unidentified sources is also high: 3.0$\pm$2.3.
When considered separately, BL LACs seem to be less variable than
quasars, a trend first noticed by Mukherjee et al. (1997) --see
also Nolan et al. 2003.

There is no clear dependence of the variability of the sources with
latitude. However, an apparent trend of increasing the variability
status for the sources with the steepest spectra is found (Torres et
al. 2001a-b, Reimer 2001).
%is shown in this figure.
Nevertheless, the latter is not yet conclusive: results for a
Spearman Rank test are in the range of a few percent for this to
be a random phenomenon. Both samples look quite similar, with no
apparent deviation from one other in their variability or photon
spectral index distributions.

No correlations appear either considering the flux, i.e. $F$ vs.
$\Gamma$ or $F$ vs. $I$, but there is a clear contrast on the flux
values: whereas most 3EG catalog $\gamma$-ray AGNs have average
fluxes above $10^{-7}$ photons cm$^{-2}$ s$^{-1}$, most of the
unidentified sources have lower values.

\subsection{The multiwavelength approach for the
identification of EGRET blazars}

3EG catalog AGNs were largely selected from the Green Bank 4.85 and
1.4 GHz single dish surveys (Condon et al. 1991, White \& Becker
1992) and the 4.85 GHz Parkes-MIT-NRAO (PMN, Griffith and Wright
1993) for the southern sky. Both surveys have a threshold flux
density of $\sim 30$ mJy and a position uncertainty of $\sim 20''$,
and both are confusion-limited within 3 degrees of the Galactic
plane. The use of 5 GHz to establish potential counterparts has no
special physical meaning, rather it is used because of the existence
of a complete survey. The $\gamma$-ray/radio correlation was
discussed by several authors (both in fluxes and luminosities, e.g.,
Padovani et al. 1993, Stecker et al. 1993, Salamon \& Stecker 1994,
Dondi \& Ghisselini 1995, Stecker \& Salamon 1996, M\"ucke et al.
1997, Zhang et al. 2001, and Cheng et al.  2000). Mattox et al.
(1997) noted that, at least, all EGRET blazars with peak
$\gamma$-ray flux above $10^{-6}$ photons~cm$^{-2}$~s$^{-1}$ were
bright ($S_5 > 1$ Jy) radio sources. For dimmer ($S_5 < 500$ mJy)
radio sources, the search for associations is more difficult (e.g.,
Wallace et al. 2002, Mirabal et al. 2000, Halpern et al. 2003,
Sowards-Emmerd et al. 2003).  Should the radio/$\gamma$-ray
correlation be non-linear, but present a trend toward low
$S_5/[F(>100)$ MeV] with increasing $\gamma$-ray flux, then the
identifications of Mattox et al. (1997) must necessarily be
incomplete, since they would be lacking the low end in radio flux.
3EG J0743+5447, with $S_5=272$ mJy, is representative of a small
group of EGRET blazars that are dim and flat at 5 GHz but have been
found to have brighter and flatter spectra extending beyond 200 GHz
(Bloom et al. 1997). 3EG J2006$-$2321, recently identified with PMN
J2005$-$2310, a flat-spectrum radio quasar with a 5~GHz flux density
of 260 mJy, may be another source of this type (Wallace et al.
2002).

Mattox et al. (2001) searched for potential radio counterparts to
all sources listed in the 3EG, allowing for sources (up to the
extent of their catalogs) with arbitrarily low $S_5$. They list 46
blazar identifications with a `high probability' of being correct,
and 37 additional `plausible' radio associations (including 15
unidentified sources, none at high latitudes).  As noted by
Wallace et al. (2002), the fact that there is no unidentified
$\gamma$-ray source at high latitude at least plausibly associated
with a blazar under this identification method seems to indicate
it might lack identification power. However, four
`high-probability' blazars were reported with $S_5<1$ Jy.

Sowards-Emmerd et al. (2003) additionally used the 3.5~cm CLASS
survey (Myers et al. 2002) to search for counterparts. The CLASS
survey targeted compact gravitational lens candidates,
preselecting flat spectrum sources by comparing the NVSS (21 cm)
and Green Bank (6 cm) fluxes. CLASS observed sources lying at DEC
$ \ge 0^{{\rm o}}$ and $|b| \ge 10^{{\rm o}}$, and having spectral
index $\alpha \le 0.5$ ($S_\nu \propto \nu^{-\alpha}$) and
resolved structure up to sub-arcsecond scales. Such resolution
showed that some flat sources were actually produced by extended
high frequency emission. Sowards-Emmerd et al. introduced a new
figure of merit to evaluate the plausibility of counterparts
defined as
%\be
${\rm FoM} = n_{8.4\; {\rm GHz}} \times n_{\alpha} \times n_{{\rm
X-ray}} \times L(\alpha,\delta).
%\ee
$ Each $n$-value computes the over-density of sources near
high-latitude ($| b | > 20^{{\rm o}}$) $\gamma$-ray detections in
bins of radio flux, spectral index, and X-ray flux, respectively.
The last factor corresponds to the source position weighting,
given by the value of the $\gamma$-ray likelihood at the radio
source position, extracted from EGRET maps. Sources with FoM $>1$
are designated as `likely' counterparts and those having $0.25 <
{\rm FoM} < 1$ are considered `plausible' counterparts.
% The results of these classification are shown in Fig. \ref{sow2}.
Out of 116 Northern 3EG catalog sources (excluding the Solar
flare), 66 have at least one plausible blazar-like radio
counterpart within this new scheme. Noteworthy, this method
proposes 50\% more high-confidence classifications than Mattox et
al. (2001), with nearly twice maximum redshift. Several
identifications are proposed with low (well below 1~Jy) radio
fluxes.

If the latter identification scheme proves to be valid,  some of
the population studies using both the sample of A-AGN out of the
3EG catalog and the 46 high-confidence AGNs of Mattox et al.
(2001) could have serious problems. Nevertheless, it is expected
that multiwavelength correlations using these samples can still be
useful to guide the forthcoming research (see, e.g., Cheng et al.
2000, and Mei et al. 2002).

Bloom et al. (1997) noted that blazars thought to have been detected
by EGRET often have flatter spectra from 5 to 22 GHz than blazars
that were not. Higher radio frequency observations (90 and 230 GHz)
of 12 southern AGNs classified as possible EGRET identifications
were made by Tornikokski et al. (2002). They found that several are
blazars, and confirmed that the AGN identifications in the 3EG
catalog are objects that are bright and variable in the mm domain,
having a flat spectrum up to 100 GHz. Optical microvariability
(Romero et al. 2002) was searched for a sample of 20 southern EGRET
AGNs, and timescales of variation of the order of several hours were
found. These results place some doubt on earlier claims (e.g., Dai
et al. 2001) on the existence of tens-of-minutes optical variability
timescales.

%VLBA

von Montigny et al. (1995a), among other authors, reported that
EGRET tended to detect superluminal radio sources. Recently,
Jorstad et al. (2001a) completed an extensive VLBA monitoring
program of 42 presumed $\gamma$-ray bright blazars, finding
apparent superluminal jet velocities in 33
sources.\footnote{Earlier studies on superluminal blazars that
were {\it not} detected by EGRET were reported by von Montigny et
al. (1995b)}. Jorstad et al. (2001b) concluded, from the relative
timing of superluminal ejections and $\gamma$-ray flares, that
superluminal phenomena and $\gamma$-ray flares are correlated.
%(statistical
%simulations show that if the number of coincidences is bigger than
%10, as it seems to be the case, the radio and $\gamma$-ray events
%are associated with each other at greater than 99.999\%
%confidence).
The population of bright $\gamma$-ray blazars detected by EGRET
can therefore be categorized as highly superluminal, with apparent
speeds as high as $\sim 40c$ (for a Hubble constant of 65 km
s$^{-1}$ Mpc$^{-1}$); the peak of the distribution being at 8--9
h$^{-1} c$, significantly higher than the average speed of jet
components in the general population of strong compact radio
sources.

% COMPTEL

The COMPTEL experiment (0.75--30 MeV) detected 11 AGNs: 9 FSRQs, 1
BL Lac (Mrk 421), and the radio galaxy Centaurus A (see e.g.,
Sch\"onfelder et al. 2000,  Collmar 2002). Upper limits were
obtained for several tens of presumed EGRET AGNs and unidentified
high-latitude EGRET detections. COMPTEL detected also a handful of
unidentified high-latitude $\gamma$-ray sources itself.
%(including GRO
%J1040+48 GRO, J1214+06 and the extended source GRO J1753+57).
%No Seyfert galaxy was detected, despite searches (Maisack et al. 1995).
COMPTEL AGNs were often visible during flaring episodes that were
simultaneously detected by EGRET, and many, at threshold level in
at least one channel, limiting the knowledge of their spectra. For
those observations that admitted a spectral fit, power-laws with
slopes $\sim$2 were found. When a simultaneous flaring detection
in EGRET was observed, COMPTEL usually saw a hardened spectrum,
showing that the upper end of the COMPTEL band was being affected
by the EGRET-detected phenomenon. COMPTEL also detected
MeV-variability of AGNs. The shortest timescale was seen for 3C279
(the most observed COMPTEL AGN), whose flux changed by a factor of
$\sim 4$ within a period of 10 days correlated with the large
EGRET flare of early 1996 (Collmar et al. 1997). In any case, no
COMPTEL AGN was detected in all corresponding pointings, implying,
too, time variation in their MeV emission. COMPTEL (together with
OSSE and EGRET) data have shown that there is a spectral turnover
in the MeV band, e.g., that is clearly seen in the already
commented cases of PKS 0528+134 and 3C279, for which it was shown
that the luminosity across the electromagnetic spectrum peaks near
the COMPTEL band.
%The MeV spectrum of Cen A is softer
%than the typical blazar at this energy band (see below the
%section on radio galaxies).

Of special interest to the INTEGRAL mission are the MeV blazars
(AGNs that are exceptionally bright at MeV energies and do not
present significant emission in the GeV band). The first detection
of a MeV blazar was found by COMPTEL (GRO J0516$-$609, Bloemen et
al. 1995) and quickly, other such object was reported  (PKS
0208$-$512, Blom et al. 1995). Different theoretical models to
explain these objects are discussed by Bednarek (1999), see also
Romero (1996).

%TeV

At higher energies, the class of EGRET blazars has been thoroughly
observed by \v{C}erenkov telescopes, but only a few have been
detected  (including the already mentioned Mkn 421, Mkn 501, PKS
2155-304). Mrk 421 constituted the first detection of $\gamma$-ray
emission from a BL Lac object (Lin et al. 1992). Mkn 501, instead,
was initially not an EGRET source, but by raising the energy
threshold and looking for flares, Kataoka et al. (1999) were able
to report its detection. A power-law fit to the EGRET data
suggested a spectral index of $1.6\pm 0.5$ for Mrk 501; the
hardest known blazar spectrum at GeV energies. EGRET spectra
extrapolation
%1ES2344+514,
down to TeV-energies are, however, consistent with non-detection
by current \v{C}erenkov telescopes.

\subsection{Are all unidentified high-latitude $\gamma$-ray sources
AGNs?}

Whichever the classification scheme adopted to identify EGRET
sources at high-latitudes, they fail to associate all unidentified
detections with AGNs. For instance, out of Mattox et al.'s (2001)
sample, only one source not classified by Hartman et al. (1999) as
A-AGNs was suggested as a high confidence AGN, leaving all
unidentified sources out of this category.  In the work by
Sowards-Emmerd et al. (2003), the technique used also selects
individual sources that are most likely  non-blazars: 28 previously
unidentified sources are so selected, some of which are located at
high latitudes.

\begin{figure*}[t]
\begin{center}
\includegraphics[width=4cm,height=5cm]{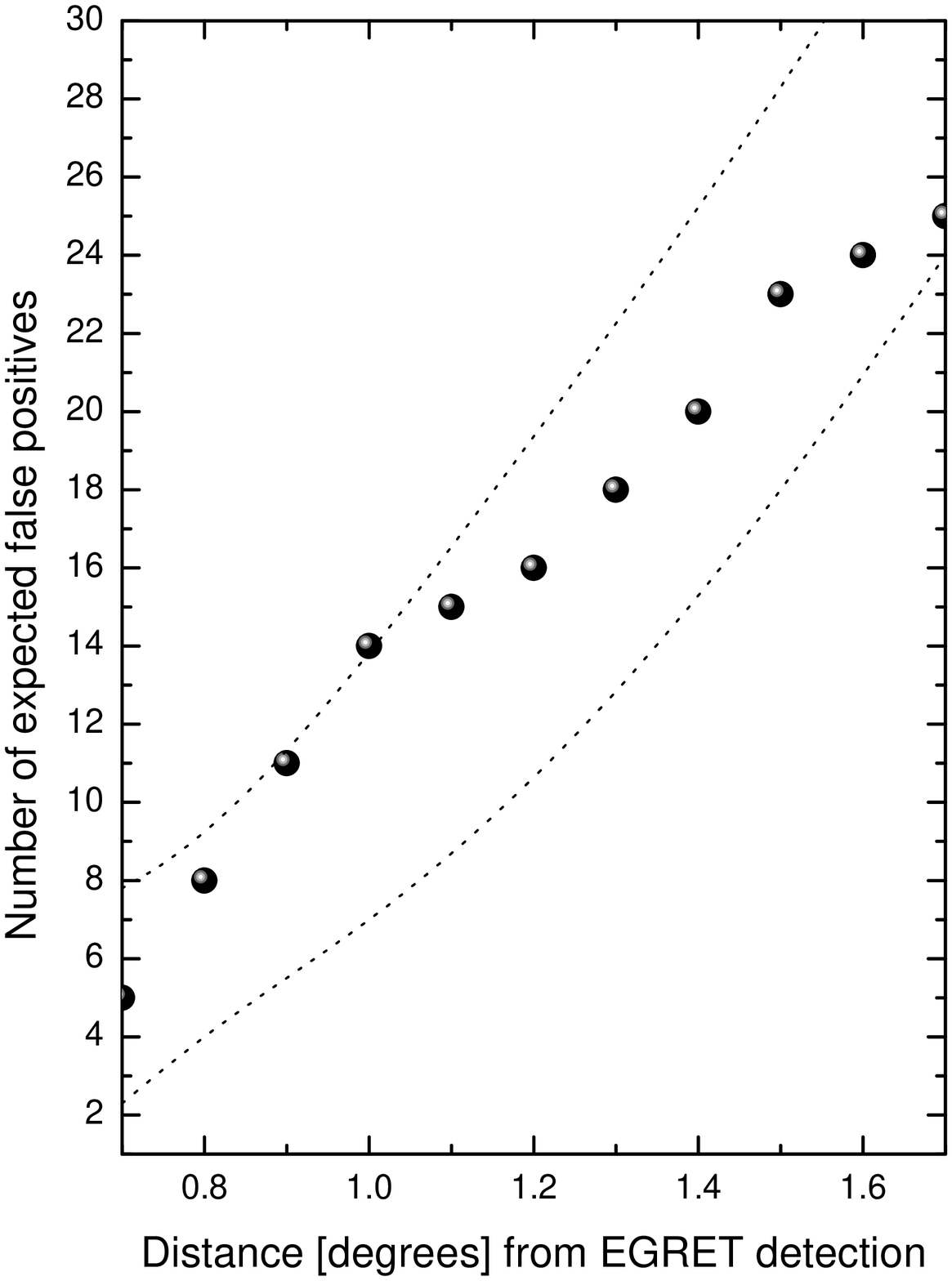} \hspace{1cm}
\includegraphics[width=5cm,height=5.15cm]{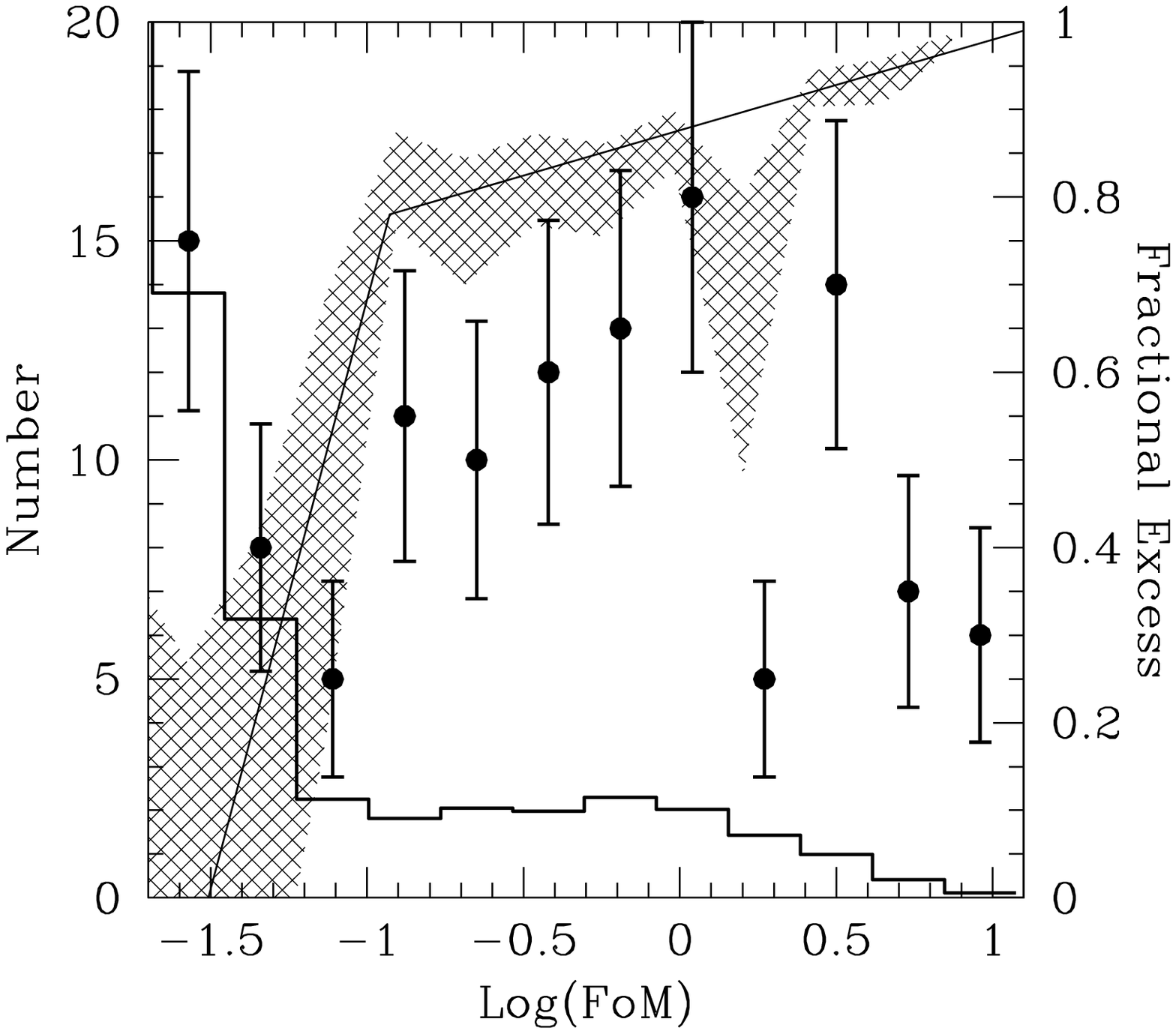}
\caption{Left: The expected distribution of radio-loud quasars
(louder than 0.5 Jy at 5 GHz) to occur by random chance as a
function of the distance from the center of the field for a sample
of 114 EGRET detections. Points represent the number of
$\gamma$-ray detections for which the counterparts are beyond the
95\% confidence contour. The dotted curve are the boundaries of
the 68\% confidence band for the hypothesis that the radio sources
are randomly distributed in the EGRET detection fields. Adapted
from Punsly (1997). The number of sources whose possible
counterpart are beyond the 95\% confidence contour is compatible
with the chance expectation. Right: Random (histogram) and true
(points, with Poisson error bars) distributions of the FoM. The
fractional excess (true ID fraction) for each bin is shown by the
(Poisson) error range of the shaded region (right scale). After
Sowards-Emmers et al. (2003).} \label{sow}
\end{center}
\end{figure*}

Furthermore, some of the associated AGNs are probably false
positives (i.e. AGNs that are mis-associated with EGRET sources by
failure of the statistical methods used in the classification). This
fact is particularly important for statistical methods based {\it
only} on the relative positions between the candidate and the EGRET
source center. If the confidence contours have any significance at
all, a source should appear beyond the 95\% contour only a few
percent of the time (Punsly 1997). Working with 114 sources above
$|b|>10^{{\rm o}}$, Punsly have estimated the number of random
coincidences as a function of the field radius: $\sim 2$ (10)
quasars with more than 1 Jy of 5 GHz flux are expected to correlate
by random chance if the size of the typical EGRET angular
uncertainty is 0.7$^{{\rm o}}$ (1.7$^{{\rm o}}$). This sheds some
doubt on several of the `plausible' correlations that occur beyond
the 95\% location contours of EGRET sources, which are associated
only by position (see Fig. \ref{sow}, left panel). The significance
of the Sowards-Emmers' et al. (2003) FoM statistic is evaluated by
shifting the radio sources positions, allowing for an average FoM
distribution to be computed and compared with the true sources
result. Fig. \ref{sow} (right panel) shows these results. The hashed
region (right scale) shows the $\pm 1\sigma$ range for the estimate
of the fraction of sources in a given FoM bin that are in excess of
random counts. The number of false positives using the FoM
statistics is expected to be low: of the 35 likely sources so
selected, less than 3 false positives are expected, whereas out of
the 32 plausible sources, the number of false positives is expected
to be less than 6. In any case, any number of false positives
implies that different physical interpretation for the origin of
some EGRET sources lying at high latitude should be sought.

The evolution and luminosity function of the EGRET blazars was
used to estimate the contribution of similar unresolved objects to
the diffuse extragalactic background (Chiang \& Mukherjee 1998).
It was found that no more than $\sim 25$\% of the extragalactic
background can be ascribed to unresolved sources with jets close
to the line of sight. Additionally, only $\sim 20$--40\% of the
diffuse extragalactic emission can be attributed to unresolved
$\gamma$-ray emitting BL LACs or FSQRs located up to $z\sim 3$
(M\"ucke \& Pohl 2000). Even integrating up to $z \sim 5$,
unresolved AGNs underproduce the intensity of the extragalactic
$\gamma$-ray background, probably by more than 20\%. Previous
results (e.g., Stecker \& Salamon 1996) assumed a linear
correlation between the measured radio and $\gamma$-ray fluxes and
obtained a much larger contribution to the $\gamma$-ray
background. However, this correlation is at least a noisy one, as
discussed above.

In a sense, the search for origins of $\gamma$-ray sources at high
latitudes, other than AGNs, mirrors the efforts to distinguish the
origin of  those low-latitude Galactic sources that display
transient behavior but which cannot be associated with radio
quasars. Examples of those are 3EG J1837-0606 (Tavani et al 1997)
and 3EG J0241+6103 (Kniffen et al. 1997, Tavani et al. 1998).
Whatever objects these are (e.g., Romero et al. 1999, 2001, 2003,
Kaufman-Bernad\'o et al. 2002, Paredes et al. 2000, Punsly et al.
2001), perhaps some of them could also be found at high Galactic
latitudes.

\section{Microlensing of $\gamma$-ray blazars}

Notwithstanding the discussion of the previous section, AGNs could
also be related to some of the $\gamma$-ray detections by the
enhancement of the radiation they produce in an event extrinsic to
the source itself: in a microlensing event (see Combi and Romero
1998, Torres et al. 2002a, 2003a).

Gravitational light deflection is an achromatic phenomenon: i.e.
the deflection angle does not depend on the energy of the photon.
However, it is nevertheless possible to have  chromaticity effects
when the size of the source is dependent on wavelength.
%A large source is typically less affected by a
%microlensing magnification than a small source (e.g., Wambsganss \&
%Paczy\'nski 1991).
It was mentioned before that the size of the $\gamma$-spheres
depends on both the energy of the $\gamma$-ray photons and the
soft photon flux. For an isotropic, power-law, central source of
soft photons scattered by free electrons,
%in a warped disk,
Blandford \& Levinson (1995) obtain: $ r_{\gamma }(E)\propto E^p,
$ with $p$ depending on the details of the central source,
typically: $p\sim 1$--2.
%The larger $\gamma$-spheres are those for
%the higher photon energies.

%\subsection{Magnification maps}

The parameters that describe a microlensing scenario are the
dimensionless surface mass density $\kappa$  and the external
shear $\gamma$ (e.g., Kayser et al. 1986; Schneider \& Weiss
1987). In ray-shooting simulations (Wambsganss 1999), a large
number of light rays (of order $10^9$) are followed backwards,
from the observer to the distant source plane, through the field
of point lenses.\footnote{The minimal number of lenses that have
to be considered depends on the focusing and shear values, as well
as on the ratio between the diffuse and the total flux. The
diffuse flux ($\epsilon$) is that coming from rays that are
deflected into the receiving area from stars far outside the
region where microlenses are considered, and should be
consistently low. An approximated expression for the number of
lenses to be included in each magnification map is (Wambsganss
1999): $ N_* \sim {3\kappa^2}/[{(1-\kappa)^2-\gamma^2}] 1/
\epsilon $ which entails values from several hundreds (for $\kappa
< 0.4$) up to several hundred thousands (for $\kappa \sim 1$)
stars, in the case of zero shear and $\epsilon = 0.01$. } In the
lens plane, the deflections from individual lenses are then
superposed for each ray and these are then followed to the source
plane. There, they are collected in small pixels. The number of
rays per pixel (on average $\sim 100$ for a region typically of
$2500 \times 2500$ pixels) is proportional to the magnification at
that position. A two-dimensional map of the ray density ---a
magnification or caustic pattern--- is then produced. Sharp lines
correspond to locations of very high magnification, i.e. the
caustics. Source, lens, and observer are moving relative to each
other; this produces a variable magnification as a function of
time. When a source crosses a caustic, formally two very bright
new (micro) images appear or disappear. However, their angular
separation is much smaller than the resolution of any telescope
and only the combined total brightness is measured, producing
dramatic jumps in the observed flux.

\begin{figure*}[t]
\begin{center}
\includegraphics[width=10cm,height=11.8cm,angle=-90]{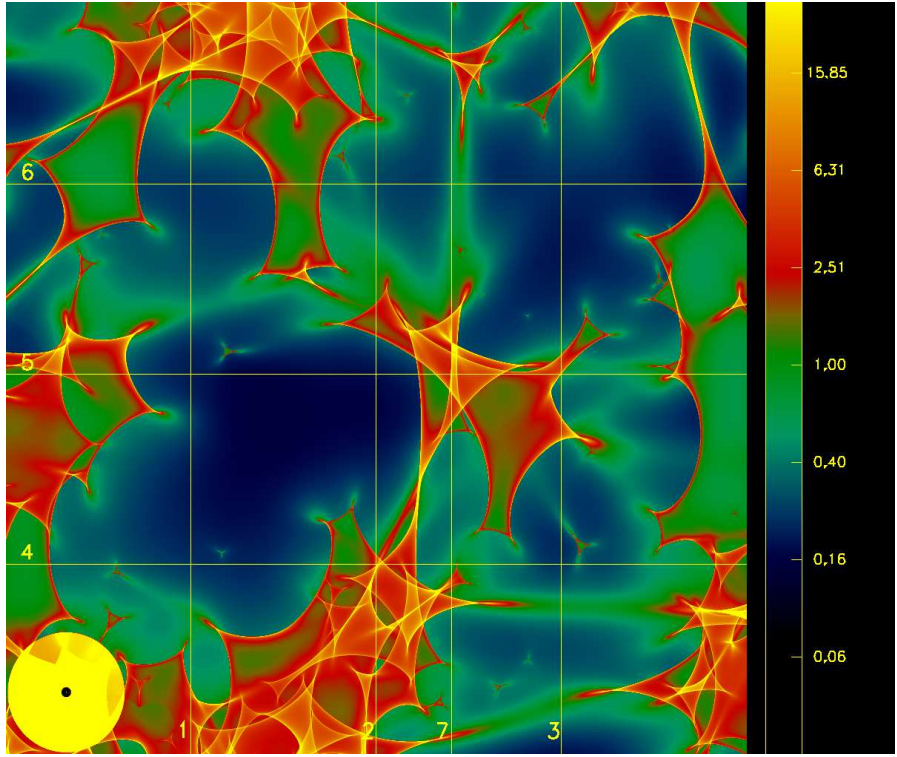}
\end{center}
\caption{Magnification map for lensing with parameters
$\kappa=0.5$ and $\gamma=0.0$: the brighter the region, the
stronger the magnification. In the bottom left of the panel, the
size of the source for three different energies is shown. The
innermost pixel (just the central point of the circles) represent
the size of the lowest energy $\gamma$-sphere, corresponding to
$E=100$ MeV. The first circle is the size of the 1 GeV
$\gamma$-sphere, while the largest circle is the size of the
$E=10$ GeV $\gamma$-ray sphere. The side length is
    10 Einstein radii of 1 M$_\odot$--star, which means the separation
    between the horizontal/vertical lines 1, 2, 3 and
    4, 5, 6 corresponds to 2.5 Einstein radii, respectively.
After Torres et al. (2003).} \label{mapk50g0}
\end{figure*}

The magnification map for the case  $\kappa=0.5$ and $\gamma=0.0$
is shown in Fig. \ref{mapk50g0}.
%The characteristic critical lines
%of the Chang-Refsdal model (1984) appear in this map.
The $\gamma$-spheres will be affected according to their size
while moving in the caustic pattern.
\begin{figure*}[t]
\begin{center}
\includegraphics[width=10cm,height=10cm]{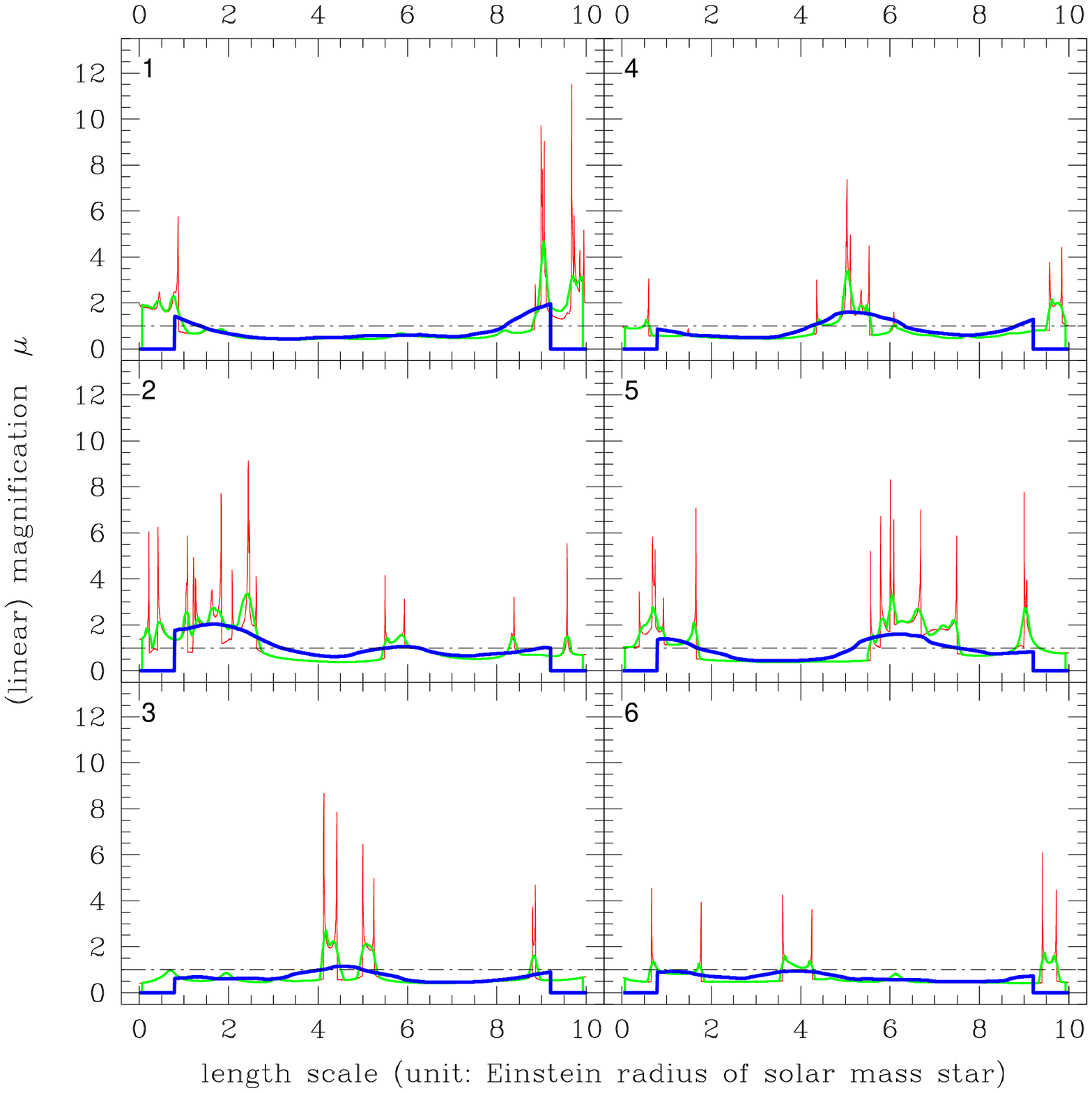}
\end{center}
\caption{Light curves for different source trajectories. Numbers
corresponds to those given in Fig.   \ref{mapk50g0}. Lines
correspond, respectively, to regions emitting photons of 100 MeV,
1, and 10 GeV (darker), and whose emitting sizes are depicted in
the bottom left corner of Fig.  \ref{mapk50g0}, the innermost
point being the less energetic $\gamma$-ray sphere. The $x$-axis
is a linear length scale, the Einstein radius of a solar mass
star, $R_E($M$_{\odot})=2.23 \;\times 10^{16}$ cm. It can be
translated into a timescale as $t=R_E($M$_{\odot})/v$, where $v$
is the relative velocity of the source with respect to the lens,
projected onto the source plane (typically of the order of 10$^3$
km s$^{-1}$). After Torres et al. (2003).} \label{curvask50g0}
\end{figure*}
The numbered lines in Fig.  \ref{mapk50g0} represent  arbitrary
source trajectories, the resulting light curves are given in Fig.
\ref{curvask50g0}. There is a typical factor of 10 more
magnification for the innermost regions than for the larger
$\gamma$-spheres (for different examples with other values of
$\kappa$ and $\gamma$, see Torres et al. 2003a).  The maximum
possible magnification ($\sim 50$) that this caustic pattern can
produce is shown by an extra trajectory that crosses exactly over
a conglomerate of several caustics (\#7).
%The magnification in
%this case can reach up to 60 times the unlensed intensity of the
%source.
Concerning timescales, the innermost $\gamma$-spheres (having
$R/{R_E} \sim 1/100$) would have a rise time of about 5 days, well
within an observing EGRET viewing period. Here, $R_E$ is the
Einstein radius $R_E=( {4GM}/{c^2}\times {D_{\rm ol} D_{\rm
ls}}/{D_{\rm os}} )^{1/2}, $ the length scale of the microlensing
problem. $D_{\rm ol}, D_{\rm os},$ and $D_{\rm ls}$ are the
angular diameter distances between the observer and the lens,
between the observer and the source, and between the lens and the
source, respectively. $G,$ $M,$ and $c$ have their usual meaning.
The largest $\gamma$-spheres, with $R\sim R_E$, can have a rise
time of about 1 yr.  Higher (lower) velocities would imply lower
(higher) timescales.

%\subsection{How many of these events could be?}

Recent results (Wyithe \& Turner 2002), taking into account the
clustering of stars in interposed galaxies, give for the a priori
probability of finding magnified sources in random directions of
the sky values between 10$^{-2}$ --- 10$^{-3}$. In those
directions where there is gravitational lensing, the probability
of having large local optical depths is high.\footnote{The concept
of optical depth was originally introduced in gravitational
microlensing studies by Ostriker and Vietri (1983).  } If the
actual total number of $\gamma$-ray emitting blazars is in excess
of 10$^7$ (1 blazar out of 10 000 normal galaxies) even when
considering reduced probabilities for microlensing, scaling as
$\tau/A^2$ with $\tau$ being the local optical depth and $A$ the
magnification, an interesting number of GLAST detections (from a
handful to some tens) could be potentially ascribed to
microlensing.

%\subsection{A plausible candidate?}

%The proposed microlensing scenario could, perhaps, be responsible
%for the $\gamma$-ray variability of 3EG J1832-2110, which has been
%associated with PKS 1830-211 (see Combi \& Romero 1998).
%They found
%that assuming a redshift $z_s = 1$ for the background source and
%$z_l \sim 0.89$ for the lens, a sub-solar object with $M \sim 0.02
%M_\odot$ moving with a velocity of $v\sim 1000$ km s$^{-1}$ would
%be enough to produce the observed variability.

\section{Alternative origins of high-latitude $\gamma$-ray
sources}

\subsection{Galaxy clusters}

Very recently, an interest in galaxy clusters as possible EGRET
counterparts has been sparked by strong claims of positive
correlations between unidentified $\gamma$-ray sources and
$\gamma$-ray excesses and the position of clusters in the Abell
Catalog (Colafrancesco 2001, Kawasaki \& Totani 2002, Scharf \&
Mukherjee 2002). In the next chapter O. Reimer analyzes in detail
the possibility of such association.

There are several reasons to expect that galaxy clusters emit
$\gamma$-rays. Hadronically produced $\gamma$-rays, via $pp$
interactions of high-energy cosmic rays with the intracluster
medium (Berezinsky et al. 1997), or as the origin of a secondary
population of relativistic electrons (Atoyan \& V\"olk 2000) have
been considered. $\gamma$-ray radiation could also arise as a
result of large-scale cosmological structure formation (Dar \&
Shaviv 1995, Colafrancesco \& Blasi 1998, Waxman \& Loeb 2000,
Totani \& Kitayama 2000; Kawasaki \& Totani 2002, Miniati 2002,
Berrington \& Dermer 2002). Reimer et al. (2003) analyzed Phase 1
to 9 of EGRET observations of a sample of 58 X-ray bright galaxy
clusters with $z<0.14$. The sample selection assumed a plausible
assumption, i.e. that the brightest and nearest clusters already
detected in X-rays should be the most likely candidates to emit
observable amounts of $\gamma$-rays (this sample includes, for
instance, all known clusters exhibiting EUV excesses, or hard
non-thermal X-ray emission). The main result of this analysis is
that no galaxy cluster in the sample has been seen by the EGRET
experiment.
%The most significant $\gamma$-ray excess is the
%cluster A3532,  a (non-)detection with 1.6$\sigma$ confidence
%level.
Interestingly, even if all observations for all clusters are added
up, this results only in an upper limit of $5.9 \times 10^{-9}$
photons cm$^{-2}$ s$^{-1}$ for an average galaxy cluster.

Upper limits for particular galaxy clusters
for which there are theoretical predictions already rule out the
models by En$\beta$lin et al. (1997) and Dar \& Shaviv (1995) and
defer for further tests those of Berrington \& Dermer (2002),
Colafrancesco \& Blasi (1998) and Miniati (2002). It is to be
noted, however, that the null results for EGRET observations of
bright X-ray clusters do not imply that such systems will not be
detected by future instruments with improved sensitivity, both
ground- and space-based.

Totani \& Kitayama (2000) have also argued for merging clusters of
galaxies (which have low X-ray brightness and thus are excluded
from the Reimer et al.'s analysis) to be possible EGRET source
counterparts.  To produce the high-energy radiation, they invoked
inverse Compton scattering of accelerated electrons with photons
of the CMB. Totani and Kitayama estimated that about 30 EGRET
sources and several thousands of future GLAST sources will be
produced by merging galaxy clusters. Berrington \& Dermer (2002)
and Colafrancesco (2002) have already presented strong objections
to the estimation of the number of sources (which would have
entailed half of the expected GLAST catalog). The main argument is
the hardness of the spectral index of injected non-thermal
electrons, for which Totani \& Kitayama obtained 2.0, assuming
that the merger shocks have large Mach numbers. Detailed estimates
of Berrington \& Dermer are between 2.2 and 2.3. Since the
energies of electrons that Compton scatter the CMB radiation to $>
100$ MeV energies exceed $\sim 200$ GeV, the steeper injection
spectrum reduces the available power in these electrons by 1--2
orders of magnitude. %, providing an essential difference between
%Berrington \& Dermer's predictions and those of Totani \& Kitayama
%(2000) (and Kawasaki \& Totani 2002).
Only a few of the
unidentified EGRET sources could then possibly be associated with
merging clusters of galaxies, and these would involve the less
frequent events involving collisions of clusters with masses near
10$^{15}$M$_{\odot}$,  hard spectral indices, or dark matter
density profiles with strong central peaks.

\subsection{Normal Galaxies}

The only normal galaxy, other than the Milky Way, detected by
EGRET is the Large Magellanic Cloud (LMC, Sreekumar et al. 1992).
It was detected using data taken along four weeks, with a flux of
$(1.9 \pm 0.4) \times 10^{-7}$ photons cm$^{-2}$ s$^{-1}$ above
100 MeV. This result had been predicted by Fichtel et al. (1991)
as the output of pion decay resulting from the interaction between
cosmic ray protons and interstellar gas, assuming galactic dynamic
balance between the expansive pressures of the cosmic rays,
magnetic fields, and kinematic motions and the gravitational
attraction of matter.

It is instructive to show how to obtain the predicted flux. One
can consider that the electron spectrum is a power-law  $N(E) dE=
KE^{-\gamma} dE$, with $N(E)$ being the number of electrons per
unit energy per unit volume, and $K$ the spectrum normalization.
The intensity of the synchrotron radiation in the presence of
random magnetic fields is
\begin{eqnarray}  I_\nu=1.35 \times 10^{-22} a(\gamma) L\, K\,
B^{(\gamma+1)/2} \times \left(\frac{6.26 \times 10^{18}}{ \nu}
\right)^{(\gamma-1)/2}  \nonumber \\ {\rm erg\; cm}^{-2} \; {\rm
s}^{-1} \; {\rm sr}^{-1}\; {\rm Hz^{-1} }, \label{1}
\end{eqnarray} (Ginzburg \& Syrovatskii 1964)
where $\nu$ is the observing radio frequency in Hz, $a(\gamma)$ is
a numerical coefficient of order 0.1, $L$ is the length over which
the electrons and magnetic fields are present and $B$ is the
magnetic field strength. The normalization of the spectrum is
assumed proportional to $B^2$ both in the LMC and our Galaxy, and
the shape of the spectrum in the LMC is assumed the same as that
in the Milky Way. Then, if $K_0$ and $B_0$ are the corresponding
values of these parameters in our Galaxy, and $w(x) K_0$ is the
value in the LMC, $B=w(x)^{1/2}B_0$. Using this expression in Eq.
(\ref{1}) the scaling can be determined: \be w(x)=\left(
\frac{2.40I_\nu}{a(\gamma) L_{21} K_0} \right)^{4/\gamma+5}
B_0^{-2(\gamma+1)/(\gamma+5)} \left( \frac{\nu}{6.26 \times
10^{18}}\right)^{2(\gamma+1)/(\gamma+5)}, \ee where
$L_{21}=L/(3.09\times 10^{21}{\rm cm})$ is the distance in kpc.
Assuming best guesses for all parameters involved (see, for
instance, the Appendix of Fichtel et al. 1991), the electron
normalization can be determined. The additional assumption that
the electron-to-proton ratio is the same in the LMC as in the
Galaxy yields the proton spectrum. An estimation of the matter
column density then allows the $\gamma$-ray flux to be computed
as:
\begin{eqnarray} F(E>100 {\rm MeV}) \simeq \int d\Omega \left[ 2 \times
10^{-25} \times \frac{w(x)}{4\pi} \times \int dl (n_a +n_m)
\right]\nonumber \\ \frac 1{4\pi d^2} {\rm photons\; s}^{-1}\;
{\rm cm}^{-2},
\end{eqnarray} with $d\Omega$ being the solid angle subtended by the emitting
region, $j_\gamma=2 \times 10^{-25} \times {w(x)}/{4\pi}$ photons
s$^{-1}$ sr$^{-1}$ H-atom$^{-1}$ being the $\gamma$-ray
production, and $ \int dl (n_a +n_m)$, with $n_a$ and $n_m$ the
atomic and molecular density, respectively, being the column
density. Note that the prediction allows different emission level
contours to be plotted, depending on the position in the galaxy.
However, in order to make a direct comparison with EGRET or any
other experiment, the predicted $\gamma$-ray intensity has to be
compared with the corresponding point-spread function. Although
the predicted intensity based on the dynamic balance is in good
agreement with the EGRET result, it is also in agreement with the
cosmic ray density being the same throughout the galaxy, as if,
for instance, the cosmic ray density is  universal in origin
(e.g., Brecher and Burbidge 1972)

The LMC EGRET detection did not allowed to distinguish between
these two possibilities, although an ever increasing amount of
evidence favored the Galactic hypothesis. It was the {\it
non-detection} of the Small Magellanic Cloud (SMC) what would
settle this issue. There were three separate EGRET observations of
the SMC amounting several weeks of off-axis SMC pointings. No
significant deviation was found in the SMC region, implying an
upper limit of $0.5\times 10^{-7}$ photons cm$^{-2}$ s$^{-1}$
(Sreekumar et al. 1993). If the cosmic ray density in the SMC were
as high as it is in our galaxy, the intensity in $\gamma$-rays
would be $\sim 2.4 \times 10^{-7}$ photons cm$^{-2}$ s$^{-1}$, a
level incompatible with the experimental result. Therefore, the
distribution of cosmic rays can not be universal.
%, EGRET dixit.

A feeling of what future instruments can do by observing the SMC
can be obtained by looking at the different assumptions for
Galactic origins of the cosmic ray flux. As done for the LMC,
Sreekumar and Fichtel (1991) have previously computed the
$\gamma$-ray emission expected from the SMC, showing that there
was a strong disagreement in the magnitude of the cosmic ray
density, and then of the $\gamma$-ray flux, between that obtained
using the matter distribution and the assumption of dynamic
balance and that deduced from synchrotron radiation. Unlike the
LMC, the SMC is not expected to be in equilibrium, but in a
disrupted state. If that is the case, the synchrotron radiation
estimation is more reliable. An instrument like GLAST, then, will
shed light on the issue of the dynamical state of the SMC, by
comparing different predictions for the cosmic ray density.

It is to be noted that a significant contribution to the diffuse
flux can be due as the sum of unresolved point like sources, e.g.,
pulsars. Numerical simulations have been done both within the
polar cap (Hartmann et al. 1993) and outer gap (Zhang \& Cheng
1998) models for $\gamma$-ray emission, and the results are
comparable. In the case of the outer gap model applied to the LMC
EGRET observations, the ratios of expected fluxes to those
observed in 100, 500 and 1000 MeV are 7.5\%, 30\%, and 95\%,
respectively, showing that unresolved sources will provide most of
the flux at high $\gamma$-ray energies, something that must be
taken into account when interpreting the high energy radiation.

Very recently, Pavlidou and Fields (2001) presented an
observability study for several of the local group galaxies,
assuming that the $\gamma$-ray flux above 100 MeV is represented
by \be  F(E>100 {\rm MeV})= 2.3 \times 10^{-8} f_G\left( \frac
{\Sigma}{10^4 {\rm M}_{\odot} {\rm kpc} ^{-2}}\right) {\rm photons
\, cm}^{-2} \, {\rm s}^{-1}, \ee with $f_g$ being the ratio
between the galaxy $G$ and the Milky Way supernova rates, and
$\Sigma$ the gas mass-to-distance squared ratio. This amounts to
the assumption that supernova remnants alone are the source of
cosmic rays, and that once produced, their propagation is
described by a leaky box model, with the additional supposition of
an equal time/length of escape to that of our Galaxy. This
approach is far simpler than that followed by Fichter, Sreekumar
and coworkers when analyzing the LMC and SMC cases, although
probably not quite correct for those galaxies which are different
from ours, like the SMC. \footnote{Using the same approach,
Pavlidou and Fields (2002) have presented a computation of the
contribution to the $\gamma$-ray background produced by cosmic-ray
interactions with diffuse gas of normal galaxies. They found that
a multi-component fit (e.g., blazars + normal galaxies) of the
extragalactic $\gamma$-ray background emission is better than the
one obtained with unresolved active nuclei alone.}
Notwithstanding, their results do not differ much from Fichtel \&
Sreekumar's (1991), and are compatible with current constraints
(Sreekumar et al. 1994). For example, the Andromeda galaxy M31, a
case studied previously by \"Ozel and Berkhuijsen (1987), presents
a flux of $1 \times 10^{-8}$ photons cm$^{-2}$ s$^{-1}$,
consistent with the observational upper limit set by Blom et al.
(1999) using more recent EGRET data. This flux could be detected
by GLAST in the first 2 years of its all-sky survey with
14$\sigma$ significance. If such is the case, it will be possible
to study the correlation between regions of higher column density
and higher $\gamma$-ray emission. It could even be possible to
observe effects of the magnetic torus (e.g., Beck et al. 1996) and
the star forming ring (e.g., Pagani et al. 1999), a morphological
feature analogous to the Milky Way's H$_2$ ring extending in
radius from 4 to 8 kpc (e.g., Bronfman et al. 1988), which has
been detected in $\gamma$-ray surveys (Stecker et al. 1975). Other
results for Local Group Galaxies show that, unless the assumptions
are severely misrepresenting the physics, only M33 might have some
chance of being detected by future instruments (Digel et al.
2000).

\subsection{Starburst Galaxies}

Starburst galaxies are subject to tremendous rates of star
formation and, consequently, of supernova explosions. As such, it
is expected that starbursts should have a $\gamma$-ray flux
significantly higher than normal galaxies. The EGRET experiment,
however, did not detect any starburst, but rather put upper limits
on a handful of them. Such is the case of M82, with $F(E>100 {\rm
MeV}) < 4.4 \times 10^{-8}$ photons cm$^{-2}$ s$^{-1}$, and NGC
253, with $F(E>100 {\rm MeV}) < 3.4 \times 10^{-8}$ photons
cm$^{-2}$ s$^{-1}$ (Blom et al. 1999), the two nearest starbursts
from Earth. These limits improved previous estimates by Sreekumar
et al. (1994), obtained with shorter exposures.

M82 in the northern hemisphere, and NGC 253 in the south, are
alike. NGC 253, at a distance of $\sim 2.5$ Mpc, has been
described as an archetypal starburst galaxy by Rieke et al.
(1980), and it has been extensively studied from radio to
$\gamma$-rays (e.g., Beck et al. 1994, Paglione et al. 1996, Ptak
1997). More than 60 individual compact radio sources have been
detected within the central 200 pc (Ulvestad et al. 1999), most of
which are supernova remnants (SNRs) with ages of only a few
hundred years. The supernova rate is estimated to be as high as
$0.2-0.3$ yr$^{-1}$, comparable to the massive star formation
rate, estimated as  $\sim 0.1$M$_\odot$ yr$^{-1}$ (Ulvestad et al.
1999, Forbes et al. 1993). The central region of this starburst is
packed with massive stars. Watson et al. (1996) have discovered
four young globular clusters near the center of NGC 253; they
alone can account for a mass well in excess of 1.5$\times 10^6
$M$_{\odot}$ (see also Keto et al. 1999). Assuming that the star
formation rate has been continuous in the central region for the
last 10$^9$ yrs, and a Salpeter IMF for 0.08--100 M$_{\odot}$,
Watson et al. (1996) find that the bolometric luminosity of NGC
253 is consistent with 1.5 $\times 10^8 $M$_{\odot}$ of young
stars. Physical, morphological, and kinematic evidence for the
existence of a galactic superwind has been found for NGC 253
(e.g., McCarthy et al. 1987, Heckman et al. 1990). This superwind
creates a cavity of hot ($\sim10^8$ K) gas, with cooling times
longer than the typical expansion timescales. As the cavity
expands, a strong shock front is formed on the contact surface
with the cool interstellar medium. Shock interactions with low and
high density clouds can produce X-ray continuum and optical line
emission, respectively, both of which have been directly observed
(McCarthy et al. 1987). The shock velocity can reach thousands of
km s$^{-1}$.

A generally similar situation applies to M82, at 3.2 Mpc. The
supernova rate in M82, for instance, may be as high as 0.3
yr$^{-1}$ (Rieke et al. 1980). The total star formation rate of
the central regions of the galaxy can be as high as $\sim 10$
M$_{\odot}$ yr$^{-1}$ (O'Connell \& Mangano 1978). The far
infrared luminosity of the region within 300 pc of the nucleus is
$\sim 4\times 10^{10}$ L$_{\odot}$ (Rieke et al. 1980). There is
$\sim 1\times 10^7$ M$_{\odot}$ of ionized gas and $\sim 2 \times
10^8$ M$_{\odot}$ of neutral gas in the IR source (Rieke et al
1980, Satyapal et al. 1997). The total dynamical mass in this
region is $\sim (1$--2) $\times 10^9$ M$_{\odot}$, of which $\sim
36$\% might be in the form of young stars (Satyapal et al. 1997).

\begin{figure}
\begin{center}
\includegraphics[angle=-90,width=10cm]{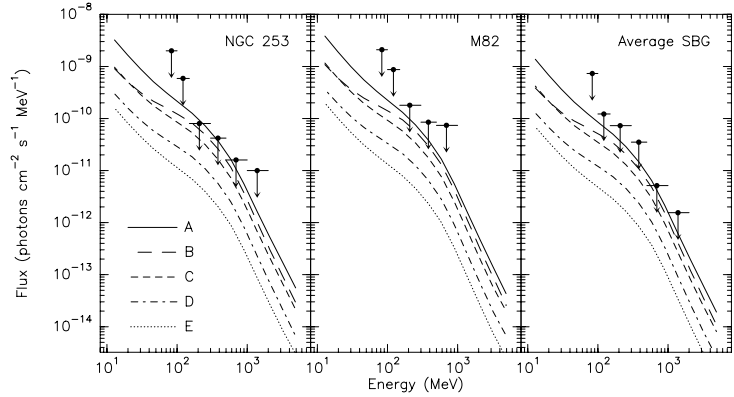}
\caption{Expected $\gamma$-ray fluxes for NGC 253, M82, and an
average starburst galaxy. Data points are EGRET upper limits
(2$\sigma$). After Blom et al. (1999).} \label{ngc}
\end{center}
\end{figure}

Early computations of the expected flux from M82 (Aky\"uz et al.
1991), have been improved recently by V\"oelk (1996), Paglione et
al. (1996) and Blom et al. (1999). Paglione et al.'s approach to
compute the $\gamma$-ray flux of a starburst is  based on a
previous study for interstellar clouds by Marscher and Brown
(1978). Results for NGC 253, M82, and that of an average starburst
galaxy [constructed co-adding observations of 10 starbursts
selected by distance ($<10$ Mpc), far infrared luminosity ($>10^9
L_\odot$) and Galactic latitude ($|b|>10^{{\rm o}}$)], are given
in Fig. \ref{ngc}.  Models with different values for the magnetic
field, the proton-to-electron ratio, and the efficiency of energy
transfer from a supernova to cosmic rays, were considered,
producing different curves in the figure.
%are given in Table \ref{tt}.

At TeV energies, at least one starburst galaxy has been detected
(NGC 253, Itoh et al. 2002). Its emission at high energy has been
interpreted as non-thermal radiation due to TeV electrons
interacting through inverse Compton scattering with the different
background photon fields (Itoh et al. 2003). The contribution of
hadronic processes in star winds was discussed by  Romero \&
Torres (2003). It is to be noted, however, that the star wind
modulation of the proton flux would inhibit protons with energies
of GeV to enter into the wind, and this process would essentially
produce a negligible output at EGRET-range energies.

%\begin{center}
%\caption{Parameter values for starburst emission models presented
%in Fig.  \ref{ngc}. After Blom et al. (1999).}
%\begin{tabular}{lcccccc}
%\hline Model & $B$ & $N_p/N_e$ & $\eta$ &
%\multicolumn{3}{c}{$F(E>100$
%MeV) ($\times 10^{-9}$ ph cm$^{-2}$ s$^{-1}$) } \\
% & ($\mu$G) & & & NGC 253 & M82 & SBGs \\
% \hline
% \hline
% A \dotfill &     25 &     50 & 0.90 & 36.5 & 37.5 & 14.4 \\ B
%\dotfill &     50 & 400 & 0.73 & 24.1 & 27.6 & 10.0 \\ C \dotfill
%&     50 & 100 & 0.43 & 15.6 & 16.7 &     6.3 \\ D \dotfill & 100
%& 100 & 0.15 &     5.3 &     6.1 &     2.2 \\ E \dotfill & 100 &
%    50 & 0.05 &     2.0 &     2.2 &     0.8 \\
%    \hline \hline
%\end{tabular}
%\end{center}
%\label{tt}
%\end{table}

The discovery of new nearby starbursts (see for example the Pico Dos
Dias Catalog, Coziol et al. 1998), enhances the probability of
$\gamma$-ray detections with GLAST, if the standard models for their
emission are basically correct (Torres et al. 2004).

\subsection{Radio galaxies}

Since the number density of radio galaxies can be a factor of
10$^3$ above those of BL LACs and FSRQs, the prospects of them
being a new population of high-latitude $\gamma$-ray sources looks
promising. Centaurus A ($l\simeq 310^{{\rm o}}, b\simeq 20^{{\rm
o}}$), at a distance of $\sim$3.5 Mpc and redshift $z=0.0018$ (Hui
et al. 1993), is the closest AGN, and the only radio galaxy
positively detected in the 3EG. It was detected with 6.5$\sigma$
confidence, appearing point-like (Sreekumar et al. 1999). The
average $>$100 MeV flux is $(13.6\pm2.5) \times 10^{-8}$ photons
cm$^{-2}$ s$^{-1}$. This photon flux implies a luminosity of
$\sim$10$^{41}$ ergs s$^{-1}$, about 10$^5$ times less than that
typical of detected blazars. This low luminosity, if typical for
radio galaxies, could explain the fact that Cen~A is the only
radio galaxy in the 3EG: more distant members of its class were
just beyond the reach of EGRET. Cen~A is identified at optical
frequencies with the galaxy NGC 5128 (Israel 1998), which has a
jet that is offset by an angle of $\sim 70^{{\rm o}}$ from the
line of sight (Bailey et al. 1986; Fujisawa et al. 2000). It
further presents a one-sided X-ray jet, collimated in the
direction of the giant radio lobes (Kraft et al. 2002, Hardcastle
et al. 2003). The radio luminosity is $\sim 10^{40}$ ergs
s$^{-1}$, and is currently classified as a Faranoff-Riley type I
radio galaxy, and earlier, as a misaligned blazar (Bailey et al.
1986). The high-energy flux of Cen~A appears to be constant, but
this could be either an intrinsic phenomenon or a biased result
due to the near-threshold detection associated with the individual
observations. The 3EG catalog 30 MeV--10  GeV photon spectrum is
well characterized by a single power law of index 2.40$\pm$0.28.
Apart from the positional coincidence, which could be suspect
because of the large localization uncertainty of EGRET, co-spatial
detections by OSSE (Kinzer et al. 1995) and COMPTEL (Steinle et
al. 1998) which provide a consistent spectrum going from 50 keV to
1 GeV, further argue for emission from single source coincident
with Cen~A. The detection of Cen~A raises the possibility that
several other unidentified $\gamma$-ray sources could be radio
galaxies.

Indeed, there are two additional EGRET sources, one of them at
high latitude, for which a possible radio galaxy counterpart has
been suggested. One such source is 3EG J1621+8203 $(l=115.5^{\rm
o}, b=31.8^{\rm o})$ (Mukherjee et al. 2002). 3EG J1621+8203
observations in individual viewing periods yielded near-threshold
detections by EGRET, as for Cen A. However, in the cumulative
exposure, it was clearly detected and the measured flux above 100
MeV was $1.1 \times 10^{ -7}$ photon cm$^{ -2}$ s${^-1}$ (Hartman
et al. 1999). The photon spectral index for this source is
2.27$\pm$0.53, steeper than the usual blazar-like spectrum.
Mukherjee et al. (2002) analyzed the X-ray and radio field
coincident with 3EG J1621+8203. They concluded that NGC 6251, a
bright Faranoff-Riley type I radio galaxy (Bicknell 1994; Urry \&
Padovani 1995) at a redshift of 0.0234 (implying a distance 91 Mpc
for $H_0= 75$ km s$^{ -1}$ Mpc$^ {-1}$), and the parent galaxy of
a radio jet making an angle of 45$^{{\rm o}}$ with the line of
sight (Sudou \& Taniguchi 2000), is the most likely counterpart of
the EGRET source. With this identification, the implied
$\gamma$-ray luminosity is also a factor of $10^{-5}$ below that
typical of blazars. Compared with Cen A, the greater distance to
NGC 6251 could, perhaps, be compensated by the smaller angle
between the jet and the line of sight.

Combi et al. (2003) have also recently reported the discovery of a
new radio galaxy, J1737$-$15, within the location error box of the
low-latitude $\gamma$-ray source  {3EG~J1735$-$1500}, whose photon
index is $\Gamma=3.24\pm0.47$. The radio galaxy morphology at
1.4~GHz is typical of the double-sided Faranoff-Riley type II. The
integrated radio flux is $55.6\pm1.5$~mJy at 1.4~GHz,  the source
is non-thermal %($\alpha\le-0.9$)
and it is not detected at 4.8~GHz
(Griffith \& Wright 1993).  Using the relation between approaching
and receding jets (e.g., Mirabel \& Rodr\'{\i}guez 1999): $S_{\rm
appr}/ {S_{\rm
rec}}=\left({1+\beta\cos\theta}/{1-\beta\cos\theta}\right)^{2-\alpha},$
as well as the radio fluxes of each jet component, a viewing angle
in the range $79^{{\rm o}}-86^{{\rm o}}$ for a velocity
$\beta=v/c$ between 0.3 and 0.9 and $\alpha=-1$ is derived.
Depending on the jet and ambient medium parameters, most
double-sided radio sources have sizes below $\sim300$~kpc
(Begelman et~al. 1984). In the case of J1737$-$15, and using
standard Friedmann-Robertson-Walker models, this size translates
into a possible distance less than 350~Mpc. If {3EG~J1735$-$1500}
is indeed the result of $\gamma$-ray emission in  {J1737$-$15},
the intrinsic luminosity at $E>100$~MeV, at the distance quoted,
should then be less than $2\times10^{44}$~erg~s$^{-1}$, also
several orders of magnitude smaller than that of blazars.

Concerning other possible radio galaxies that might have been
observed by EGRET, Cillis et al. (2003) have used an stacking
procedure to establish upper limits of the order of $10^{-8}$
photons cm$^{-2}$ s$^{-1}$. M87, a giant radio galaxy for which
there has been a recent detection of a TeV excess at a level of
4$\sigma$ (Aharonian et al. 2003), is also expected to be a source
for GLAST, having an EGRET upper limit of $2.8\times 10^{-8}$
photons cm$^{-2}$ s$^{-1}$ above 100 MeV (Reimer et al. 2003, who
improved the limit imposed by Sreekumar et al 1994), and
theoretical flux predictions not much smaller that this value
(Dermer and Rephaeli 1988).

\subsection{Cold molecular clouds in the galactic halo}

Very recently, Walker et al. (2003) have proposed that most of the
unidentified $\gamma$-ray sources at all latitudes are cold, dense
gas clouds of baryonic dark matter. These clouds would emit
$\gamma$-rays mainly by $pp$ interactions between atoms of the
cloud and cosmic rays residing in the Galactic halo.

Walker et al.'s model predicts about 300 unidentified EGRET
detections. All of these sources are expected to be extended and
non-variable, both facts at variance with current observations. It
is also expected that there will be thermal emission from the
clouds, implying bright microwave sources coincident with the
$\gamma$-ray sources. In Walker et al.'s model, the microwave and
$\gamma$-ray flux are both proportional to the same parameters (the
mass of the cloud and the cosmic-ray density). Consequently, it is
possible to scale down in frequency the $\gamma$-ray flux, in order
to obtain an estimate of the expected bolometric microwave emission.
The result is $S \sim 1.8 \times 10^{-10} F_7$ erg cm$^{-2}$
s$^{-1}$, where $F_7$ is the $\gamma$-ray flux above 100 MeV in
units of 10$^{-7}$ ph cm$^{-2}$ s$^{-1}$. The spectrum of emission
plays a crucial role: depending whether it is a blackbody or a dusty
spectrum, at low frequencies, it can be approximated by $S_\nu \sim
4.2 \nu^2 F_7$ mJy or $S_\nu \sim 1.4 \times 10^{-5} \nu^4 F_7$ mJy,
with $\nu$ in GHz, respectively. However, the Wilkinson Microwave
Anisotropy Probe data constrain this possibility.
% (WMAP, {http://map.gsfc.nasa.gov})
Of the 208 sources in the WMAP catalog, 203 sources have known
counterparts. The five without counterparts are all near the
detection threshold, and this number is compatible with the
expected number of false positives.
%The WMAP collaboration
%concluded that they are likely spurious.
Thus, there is no evidence for a population of bright microwave
sources without known radio counterparts (Bennett et al. 2003).

\section{EGRET detections and cosmic
rays}

Gorbunov et al. (2002) claimed that a set of $\gamma$-ray loud BL
Lacs can be selected by intersecting the EGRET, the ultra-high
energy cosmic ray, and the BL Lac catalogs (all conveniently cut).
The only requirement Gorbunov et al. considered for an object (a BL
Lac) to be physically associated with an EGRET source is that the
angular distance between the best estimated position of the pair
does not exceed $2\times R_{95}$, where $R_{95}$ is the 95\%
confidence level contour of the EGRET detection. This is an
unjustified assumption (see Fig. \ref{sow}, left panel), and is
contrary to previous results (Sigl et al. 2001). Torres et al.
(2003b) have searched for correlations between the same set of BL
Lacs and the arrival directions of 33 cosmic rays of ultra high
energy not used by Gorbunov et al.
%with incident zenith angle $<45^{{\rm o}}$,
%observed by the Volcano Ranch
%(6 events with energy $> 10^{19.6}$~eV)
%and the Haverah Park
%(27 events with energy $> 10^{19.4}$~eV)
%arrays.
The latter constitutes a blind sample statistically relevant to
test the hypothesis. No positional coincidences within the
accuracy of the angular determination was found.
%Taking data at face value, this implies that the measurement of
%the mean expected from the strongly correlated sample is excluded
%at the 95\% CL.
The probability that this result arises as a statistical
fluctuation from the strongly correlated sample was found to be
more than a $2\sigma$ deviation.

\section{Concluding remarks}

In this chapter, some aspects of our knowledge of high-latitude
$\gamma$-ray sources were reviewed. As during the last decade of the
previous century, we will soon enter into a period where
simultaneous multiwavelength observations will be possible. It is
expected that both a tremendous impact on the phenomenological
understanding of AGNs and the yet-unidentified high-latitude
sources, and concrete new theoretical challenges, will arise from
such forthcoming campaigns. We are on the verge of producing, once
again, an observationally driven $\gamma$-ray astrophysics.

\section*{Acknowledgments}

This work was performed under the auspices of the U.S. Department
of Energy (NNSA) by UC's LLNL under contract No. W-7405-Eng-48. I
acknowledge W. Collmar, R. Hartman, R. Romani, A. Wehrle,  J.J.
Blom, and the AAS for their kind permission to reproduce their
figures, and O. Reimer, G. Romero, S. Digel, R. Hartman, and C.
Mauche for critical readings and discussions.

{%\normallatexbib

\bibliographystyle{apalike}
\chapbblname{chapbib}
\chapbibliography{logic}

}

\end{document}